\documentclass{aastex62}

\usepackage{graphicx}   % Including figure files
\usepackage{amsmath}    % Advanced maths commands
\usepackage{bm}
\usepackage{amssymb}    % Extra maths symbols
\usepackage{diagbox}    % Extra maths symbols
\usepackage{subfigure}
\usepackage{enumerate}
\usepackage{float}
\usepackage{ulem}
\received{XXXX}
\revised{XXXX}
\accepted{XXXX}
\submitjournal{ApJ}

\shorttitle{ELM WDs in DDs}
\shortauthors{Z. Li et al.}
\begin{document}

\title{Formation of Extremely Low-Mass WDs in Double Degenerates}

\correspondingauthor{Xuefei Chen}
\email{cxf@ynao.ac.cn}

\author[0000-0002-1421-4427]{Zhenwei Li}
\affiliation{Yunnan Observatories, Chinese Academy of Sciences, Kunming, 650011, China}
\affiliation{Key Laboratory for the Structure and Evolution of Celestial Objects, Chinese Academy of Science}
\affiliation{University of the Chinese Academy of Science, Yuquan Road 19, Shijingshan Block, 100049, Beijing, China}

%\author[0000-0002-0786-7307]{Greg J. Schwarz}
%\affil{American Astronomical Society \\
%2000 Florida Ave., NW, Suite 300 \\
%Washington, DC 20009-1231, USA}

\author{Xuefei Chen}
\affiliation{Yunnan Observatories, Chinese Academy of Sciences, Kunming, 650011, China}
\affiliation{Key Laboratory for the Structure and Evolution of Celestial Objects, Chinese Academy of Science}
\affiliation{Center for Astronomical Mega-Science, Chinese Academy of Science, 20A Datun Road, Chaoyang District, Beijing 100012, China} 
\nocollaboration

\author{Hai-Liang Chen}
\affiliation{Yunnan Observatories, Chinese Academy of Sciences, Kunming, 650011, China}
\affiliation{Key Laboratory for the Structure and Evolution of Celestial Objects, Chinese Academy of Science}
\nocollaboration

\author{Zhanwen Han}
\affiliation{Yunnan Observatories, Chinese Academy of Sciences, Kunming, 650011, China}
\affiliation{Key Laboratory for the Structure and Evolution of Celestial Objects, Chinese Academy of Science}
\affiliation{Center for Astronomical Mega-Science, Chinese Academy of Science, 20A Datun Road, Chaoyang District, Beijing 100012, China} 
\nocollaboration

% Title of the paper, and the short title which is used in the headers.
% Keep the title short and informative.

% The list of authors, and the short list which is used in the headers.
% If you need two or more lines of authors, add an extra line using \newauthor

%\author{Zhenwei Li,$^{1,2,3}$
%\thanks{E-mail:lizw@ynao.ac.cn }
%Xuefei Chen$^{1,2,4}$
%Hai-Liang Chen$^{1,2}$
%Zhanwen Han$^{1,2,4}$
%Third Author$^{2,3}$
%and Fourth Author$^{3}$
% List of institutions
%$^{1}$Yunnan Observatories, Chinese Academy of Sciences, Kunming, 650011, China\\
%$^{2}$Key Laboratory for the Structure and Evolution of Celestial Objects, Chinese Academy of Science\\
%$^{3}$University of the Chinese Academy of Science, Yuquan Road 19, Shijingshan Block, 100049, Beijing, China\\
%$^{4}$Center for Astronomical Mega-Science, Chinese Academy of Science, 20A Datun Road, Chaoyang District, Beijing 100012, China 
%}

\begin{abstract}
Extremely low-mass white dwarfs (ELM WDs) are helium WDs with a mass less than $\sim$$0.3\rm\;M_\odot$. 
Most ELM WDs are found in double degenerates (DDs) in the ELM Survey led by Brown and Kilic. 
These systems are supposed to be significant gravitational-wave sources in the mHz frequency.
In this paper, we firstly analyzed the observational characteristics of ELM WDs and 
found that there are two distinct groups in the ELM WD mass and orbital period plane, indicating 
two different formation scenarios of such objects, i.e. a stable Roche lobe 
overflow channel (RL channel) and common envelope ejection channel (CE channel). We then 
systematically investigated the formation of ELM WDs in DDs by a combination of detailed 
binary evolution calculation and binary 
population synthesis. Our study shows that the majority of ELM WDs with mass less than 
$0.22\rm\;M_\odot$ are formed from the RL channel. The most common progenitor mass in this way 
is in the range of $1.15-1.45\rm\;M_\odot$ and the resulting ELM WDs have a peak around $0.18\rm\;M_\odot$ when 
selection effects are taken into account, consistent with observations. The ELM WDs with a mass 
larger than $0.22\rm\;M_\odot$ are more likely to be from the CE channel and have a peak of ELM WD 
mass around $0.25\rm\;M_\odot$ which needs to be confirmed by future observations. 
By assuming a constant star formation rate of 2$\rm\;M_\odot yr^{-1}$ for a Milky Way-like galaxy, 
the birth rate and local density are $5\times10^{-4}\rm\;yr^{-1}$ and $1500\rm\;kpc^{-3}$, 
respectively, for DDs with an ELM WD mass less than $0.25\rm\;M_\odot$. 
\end{abstract}


%%%%%%%%%%%%%%%%%%%%%%%%%%%%%%%%%%%%%%%%%%%%%%%%%%

%%%%%%%%%%%%%%%%% BODY OF PAPER %%%%%%%%%%%%%%%%%%

\section{Introduction}
\label{sec:1}

Extremely low-mass white dwarfs (ELM WDs) are helium WDs with a mass less than $\sim 0.3\rm M_\odot$. 
Recently, a number of ELM WDs and precursors have been detected by several survey projects, e.g. the Kepler project 
\citep{kerkwijk2010,carter2011,breton2012,rappaport2015}, the Wide Angle Search for Planets (WASP, 
\citealt{maxted2011,maxted2013,maxted2014,maxted2014b}), and the ELM Survey 
\citep{brown2010,kilic2011a,brown2012,kilic2012,brown2013,gianninas2015,brown2016a}. 
Up to now, the ELM Survey has discovered 82 ELM WDs in double degenerates 
(DDs\footnote{In this paper, DDs refer in particular to ELM WDs with CO WD companions, 
which are the most common systems in the ELM Survey.}, \citealt{brown2017}). 
The most compact binary \emph{J0651+2844} found in the ELM Survey has an 
orbital period of 765 s \citep{brown2011}, 
and could be a resolved source for future space-based gravitational waves detectors, such 
as the \emph{Laser Interferometer Space Antenna} (\citealt{lisa2012,brown2011,brown2017}) and 
TianQin \citep{luo2015}. 

Another interesting aspect is that many ELM WDs (or proto-He WDs, i.e. helium WD precursors) have pulsations 
\citep{maxted2011,maxted2013,maxted2014,zhangx2016,gianninas2016}, 
which are mainly driven by $\kappa-\gamma$ mechanism in the $\rm{He}^{+}-\rm{He}^{++}$ partial 
ionization zone, H-ionization region and core region \citep{jeffery2013,corsico2012,grootel2013,corsico2014a,corsico2016}, 
and some pulsations are powered by stable H burning via the $\varepsilon$ mechanism \citep{corsico2014b}. 
The pulsations allow us to study the structure of these objects in detail via asteroseismology 
\citep{calcaferro2017}. 
For example, \citet{istrate2016b} modeled the pulsating ELM WDs by considering rotational mixing 
and explained the mixed atmosphere of such objects \citep{gianninas2016}. 

Observationally, three types of companions of (proto-) ELM WDs have been discovered, that is, 
the A- or F-type dwarfs (EL CVn-type binaries, \citealt{maxted2011,maxted2013}), the 
millisecond pulsars \citep{istrate2014a, istrate2014b} and WDs, such as those in the ELM Survey. 
From the point of view of binary evolution theory, ELM WDs may be formed from either the stable 
Roche lobe overflow channel (RL channel) or the common envelope ejection channel (CE channel). However, the 
study of \citet{chen2017} shows that the CE channel cannot reproduce any of the observed EL CVn-type 
binaries due to the fact that the released of orbital energy is not enough to eject the common 
envelope (CE), which is tightly bounded when the donor is near the base of the red giant. 
Meanwhile, ELM WDs with millisecond pulsars are also unlikely to be produced by the CE channel, since the 
neutron stars (NSs) cannot accrete enough material to be a millisecond pulsars. 

In this paper, we systematically investigate the formation of ELM WDs in DDs, explain their observational 
properties, and give the birth rate and local density for further study of their contribution to the 
foreground of gravitational-wave rdiation (GWR). The observations are briefly summarized in 
Section~\ref{sec:2}, the formation channels are demonstrated in Section~\ref{sec:3}, and 
the simulation method is introduced in Section~\ref{sec:4}. Results and conclusions are 
presented in Section~\ref{sec:5} and Section~\ref{sec:6}, respectively. 

\section{Observations}
\label{sec:2}
The ELM Survey is a targeted survey project of ELM WDs by color and de-reddened $g-$band magnitudes
($15<g_0<20\;\rm mag$), operated at the 6.5m MMT telescope by \citet{brown2010}. This program started 
in 2009, and found many ELM WD samples. Following are the introduction of selection effects and the 
observed samples in the ELM WD mass - orbital period plane.  
\subsection{Selection effects}
\label{subsec:2.1}
The ELM WDs in DDs in our study are from the ELM Survey. 
To ensure the completeness of the observed samples, \citet{brown2016a} defined a `clean' sample of ELM WDs 
in the ELM Survey. First, they restricted the samples with semi-amplitude $k>75\rm\; km~s^{-1}$ and orbital period 
$P_{\rm orb}<2\rm\; days$ based on the sensitivity tests. Then they selected samples with surface gravity of 
$4.85<\log g<7.15$ and color selection of $8000<T_{\rm eff}<22000\rm\; K$ to obtain a high completeness of 
follow-up observations (see also \citealt{brown2016b}). Finally, 62 objects\footnote{\citet{brown2016b} 
removed two ELM WDs, \emph{J0345+1748} and \emph{J0308+5140}, since they are not presented in the Sloan 
Digital Sky Survey (SDSS) photometric catalog.} were selected from a total of 82 ones into the clean sample. 
The parameters for each ELM WD--i.e., the orbital periods, $P_{\rm orb}$; the ELM WD mass $M_{\rm He}$; 
and the companion mass, $M_{\rm CO}$--can be found in \citet{brown2016a}. Our theoretical studies will 
be compared with the clean sample in Section~\ref{sec:5}. In the clean sample, all of the companions of ELM 
WDs are more massive than $\sim$$0.5\;\rm M_\odot$, except for \emph{J0745+1949}, which companion could 
be another ELM WD \citep{brown2012, hermes2013}. We therefore only consider the companions to be CO WDs 
in our study. 

\begin{figure}
	\centering
	\includegraphics[width=0.6\textwidth]{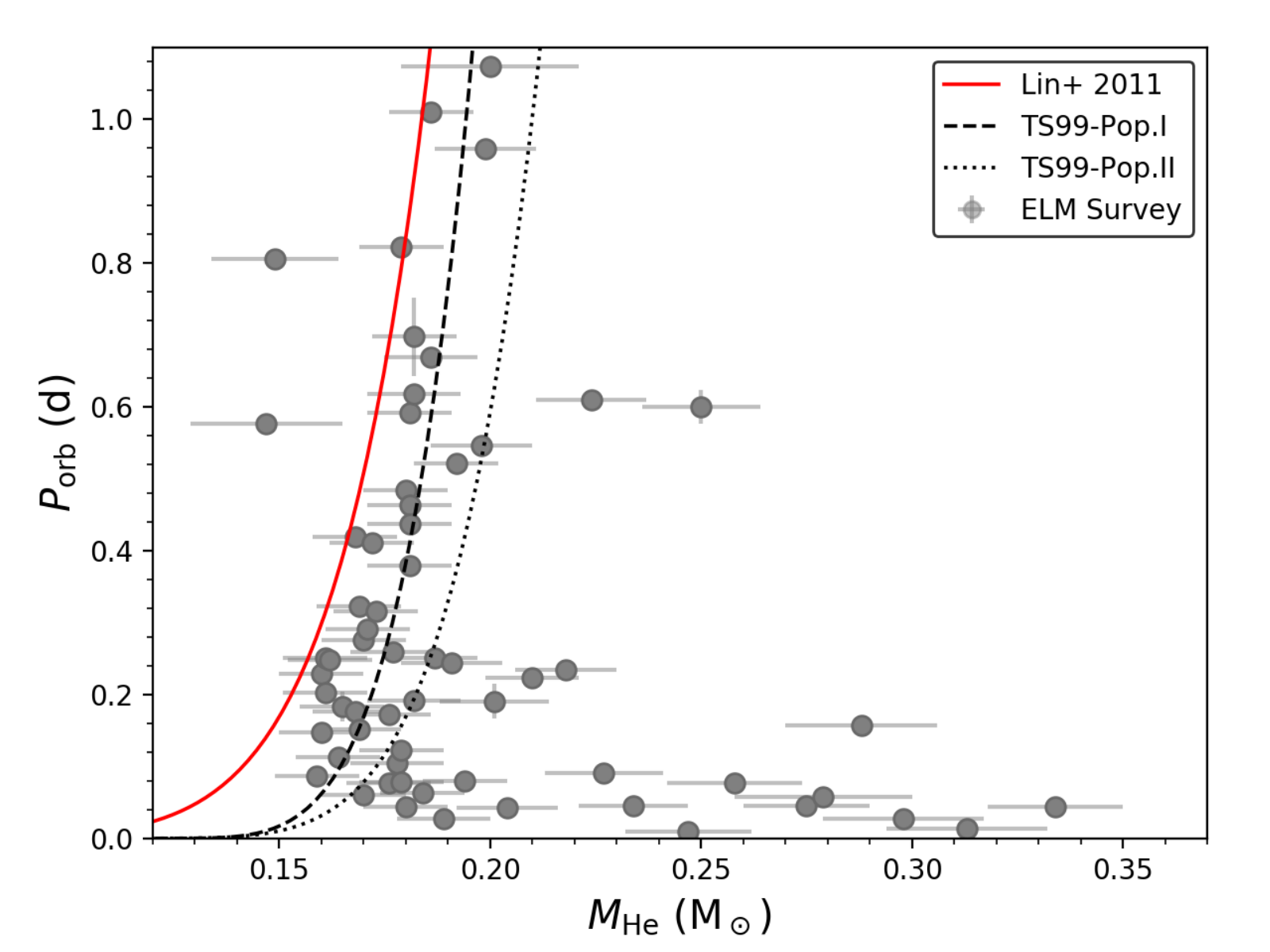}
    \caption{The ELM WD mass vs. orbital period for ELM WDs in the clean sample from 
	\citet{brown2016a}. The red solid line is from \citet{lin2011} based on detailed binary evolution 
	calculation, and the black lines are those from \citet{tauris1999} for Population I	($Z=0.02$, dashed) 
	and Population II ($Z=0.001$, dotted), respectively. 
	}
    \label{fig:1}
\end{figure}
\subsection{The ELM WD Mass - Orbital Period Plane} 
\label{subsec:2.2} 
The ELM WD mass and orbital period have some hints for the formation of ELM WDs in DDs. 
For example, both \citet{chen2017} and \citet{istrate2014b} showed a unique relation between 
$M_{\rm{He}}$ and $P_{\rm{orb}}$ for ELM WDs resulting from stable mass transfer (MT). In 
order to understand the evolutionary scenario of ELM WDs in DDs, we put the clean samples 
in the $M_{\rm{He}}-P_{\rm{orb}}$ plane and compare with some theoretical WD mass - orbital period 
relations from detailed binary evolution calculations as shown in Figure~\ref{fig:1} where the 
red solid line is from \citet{lin2011} and the black dashed and dotted lines are from 
\citet{tauris1999} for Population I and II stars, respectively\footnote{It seems that the results of 
\citet{tauris1999} match the observations better, but their binary evolution calculations have not, in fact, 
included products with such short periods ($<1\;\rm d$). Many detailed binary evolution 
calculations with such short-period products \citep{istrate2016b,istrate2016a,chen2017} 
show relatively longer orbital periods than that of \citet{tauris1999}, similar to that of 
\citet{lin2011} shown in the figure.}. 
We see that some systems follow the theoretical $M_{\rm He}-P_{\rm orb}$ 
relation, but some have orbital periods much shorter than that derived from this relation, indicating two 
distinct formation channels for such objects, i.e. the RL channel for the former and the CE channel for the 
latter.

\section{Formation channels for ELM WDs in DDs}
\label{sec:3}
\begin{figure}
	\centering
	\includegraphics[width=0.6\textwidth]{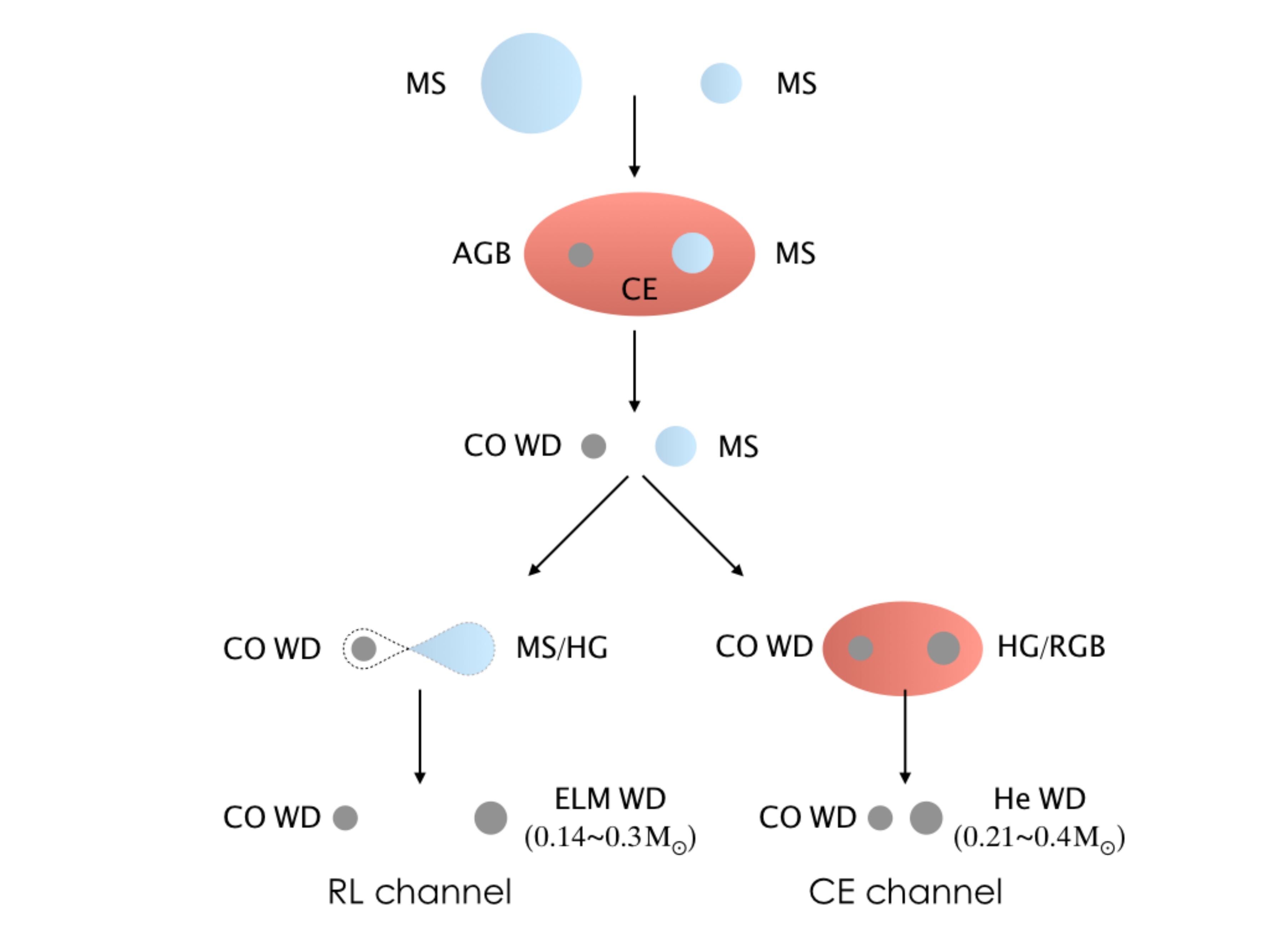}
    \caption{A sketch map for the formation of DDs with ELM WDs. 
	MS --- main sequence, AGB --- asymptotic giant branch, CE --- common envelope, 
	HG --- Hertzsprung gap, RGB --- red giant branch. 
	}
    \label{fig:2}
\end{figure}
Figure~\ref{fig:2} shows a sketch map for the formation of ELM WDs in DDs. We describe this from the 
point of view of binary evolution as below. 

(1) The primary (the initially more massive one) evolves and fills its Roche lobe when it is on the asymptotic giant 
branch (AGB), while the secondary is still on the main sequence (MS). The MT process is dynamically unstable 
(depending on the mass ratio), and the binary enters into a CE evolution phase. A 
CO WD + MS system is formed after the ejection of the CE. Stable MT from an AGB star to an MS 
companion may also result in a CO WD + MS system but with a very long orbital period, and it cannot contribute 
to the formation of DDs with ELM WDs. If the primary fills its Roche lobe before it becomes an AGB star, 
the product may be an He star + MS system (either from stable MT or CE evolution), which may further evolve into 
a CO WD + MS system with a relatively short orbital period and then contribute to the formation of ELM WDs in DDs. 
However, as we checked, this contribution is very small ($<0.2\%$, see also \citealt{willems2004}). 
We thus have not shown this in the sketch map for clarity. 

(2) The CO WD + MS system evolves into a DD binary with an ELM WD through the RL channel or CE channel. 

\emph{RL channel}: The secondary (the MS star in the system) fills its Roche lobe in the late MS or during 
the Hertzsprung gap (HG; beyond but very close to the bifurcation point in low-mass binary evolution; see 
\citealt{chen2017}), starting to transfer mass to the CO WD. The MT process is stable (depending on binary 
parameters; see Section~\ref{subsec:5.1}), and a proto-He WD is formed after the termination of MT. The system 
further evolves into a DD binary, while the mass of most ELM WDs is in the range of 0.14$\sim$0.30$\;\rm M_\odot$. 

\emph{CE channel}: The secondary fills its Roche lobe during HG or near the base of the red giant branch 
(RGB). The MT is dynamically unstable, and the system enters into the CE process again. A proto-He WD 
is produced if the CE is ejected in the following evolution, and the system eventually becomes a DD binary. 
The He WD formed in this way may be as low as $\sim0.21\;\rm M_\odot$ as we show in Section~\ref{subsec:5.2}. 
Lower-mass ELM WDs cannot be produced from this channel due to the fact that the high binding energy of the CE 
in such a binary leads to the merger of the system rather than the ejection of the CE during the CE evolution. 

\section{Methods}
\label{sec:4}
To investigate the formation of ELM WDs in DDs, we firstly studied the parameter space from the RL channel 
by detailed binary evolution and then obtained the population properties from a binary population synthesis 
approach. The CE channel is included in the binary population synthesis approach. 

\subsection{Binary evolution}
\label{subsec:4.1}
\subsubsection{The binary evolution grid}
\label{subsubsec:4.1.1}
The binary evolution is done with the \emph{Modules for Experiments in 
Stellar Astrophysics} (\texttt{MESA}; \citealt{paxton2011, paxton2013, paxton2015}) code. 
We utilized the binary module in \texttt{MESA} version 9575. For convenience, the accretor is assumed to be 
a point mass. For the donor star, we adopted the element abundances of Population I stars, i.e. 
metallicity $Z$= 0.02 and hydrogen mass fraction $X$ = 0.70. 
The mixing-length parameter is set to be $\alpha_{\rm{MLT}}=1.9$. The MT rate is given by 
the scheme of \citet{ritter1988}, that is, 
\begin{equation}
	\dot{M} \propto \frac{R^3_{\rm{RL,d}}}{M_{\rm{d}}}\exp\left(\frac{R_{\rm d}-R_{\rm{RL,d}}}{H_{\rm{p}}}\right),
	\label{eq:1}
\end{equation}
where $R_{\rm d}$ and $R_{\rm{RL,d}}$ are the stellar radius and RL radius of the donor, and 
$H_{\rm{p}}$ is the pressure scale height. We stop the evolution when the evolutionary age reaches 13.7 Gyr. 

We start our binary evolution calculations from a series of zero-age MS stars with CO WD 
companions, which are the most common companions in observations (see Section~\ref{subsec:2.1}). The mass of 
the MS star, $M_{\rm d,i}$, ranges from 1.0 to 2.0$\;\rm M_{\odot}$ in steps of 
0.1$\;\rm M_\odot$, and the CO WD, $M_{\rm CO,i}$, ranges from 0.5 to 1.1 $\rm M_\odot$ in steps of 
0.1$\;\rm M_\odot$. We also consider the case of 0.45$\;\rm M_\odot$ CO WDs, which is generally 
considered to be the minimum mass of a CO WD\footnote{\citet{han2000,chen2002,chen2003} obtained hybrid 
WDs (a CO core with a thick He shell) with masses as low as $\sim$$0.33\;\rm M_\odot$, but the structure 
of the hybrid WDs is significantly different from normal CO WDs and the accretion behaviors could be 
similar to He WDs more likely due to the thick He shells.}.

The choice of initial orbital periods depends on the bifurcation period, $P_{\rm b}$. For systems 
with initial orbital periods $P_{\rm orb,i }<P_{\rm b}$, the donors evolve to smaller masses and 
luminosities and cannot develop a compact core. So the ELM WDs can be formed from the RL channel 
only when the initial period is longer than $P_{\rm b}$ (see \citealt{chen2017} for more details). 
Meanwhile, the exact value of $P_{\rm b}$ changes with the assumptions of binary evolution. So we 
will firstly find out $P_{\rm b}$ based on our assumptions of binary evolution introduced below, 
and then we increase $P_{\rm orb,i}$ from $P_{\rm b}$ in steps of 0.02 d if $P_{\rm orb,i}<4.5\;\rm d$ 
and of 1.0 d if $P_{\rm orb,i}\geq \rm 4.5 \;d$. The upper limit of $P_{\rm orb,i}$ is for that when 
the MT rate during RL overflow (RLOF) is up to $10^{-4}\;\rm M_{\odot} yr^{-1}$ (if the MT rate exceeds 
$10^{-4}\;\rm M_{\odot} yr^{-1}$ the accretor expands rapidly, and we assume that the binary 
enters the CE phase soon after that; see more details in \citealt{chen2017}), or the He core mass is larger than 
0.4 $\rm M_\odot$. 

\subsubsection{Accretion of CO WDs}
\label{subsec:4.1.2}
The scenario for CO WD accretion is assumed to be similar to that in the study of SNe Ia 
\citep{hachisu1996,han2004}. We briefly introduce this as follows. 
There is a critical MT rate $\dot{M}_{\rm{cr}}$ for CO WDs, which is defined by 
\begin{equation}
	\dot{M}_{\rm{cr}}=5.3\times 10^{-7}\frac{(1.7-X)}{X}(M_{\rm{CO}}-0.4),
	\label{eq:2}
\end{equation}
where $X$ is the hydrogen mass fraction, and $M_{\rm CO}$ is the mass of the CO WD. If the MT rate 
$|\dot{M}_{\rm{d}}|>\dot{M}_{\rm{cr}}$, the accreted hydrogen burns steadily on the WD surface 
with a mass accumulation rate $\dot{M}_{\rm{cr}}$. The unprocessed matter 
is assumed to be lost in the form of optically thick wind at a rate of 
$\dot{M}_{\rm{wind}}=|\dot{M}_{\rm{d}}|-\dot{M}_{\rm{cr}}$ \citep{hachisu1996}. 
There is no mass loss (ML) and hydrogen burning is steady when 
$\frac 1 2\dot{M}_{\rm cr}<|\dot{M}_{\rm{d}}|<\dot{M}_{\rm{cr}}$. 
For $\frac 1 8\dot{M}_{\rm cr}<|\dot{M}_{\rm d}|<\frac 1 2\dot{M}_{\rm cr}$, 
owing to the weak shell flashes caused by the unstable hydrogen-shell burning, 
it is assumed that the processed mass can be retained. 
As for $|\dot{M}_{\rm d}|<\frac 1 8\dot{M}_{\rm cr}$, strong hydrogen-shell 
flashes will eject all accreted material (see also \citealt{nomoto2007}). 

The mass growth rate of the helium layer mass under the 
hydrogen-burning shell is defined as 
\begin{equation}
	\dot{M}_{\rm{He}}=\eta_{\rm{H}}|\dot{M}_{\rm{d}}|,
	\label{eq:3}
\end{equation}
where $\eta_{\rm{H}}$ is the mass accumulation efficiency for hydrogen-burning 
(see \citealt{hachisu1999}): 
\begin{equation}
	\eta_{\rm{H}}=
	\begin{cases}
	\dot{M}_{\rm{cr}}/|\dot{M}_{\rm{d}}|,\quad\quad |\dot{M}_{\rm{d}}|>\dot{M}_{\rm{cr}} \\
	1,\qquad\qquad\quad \dot{M}_{\rm{cr}}\ge|\dot{M}_{\rm{d}}|\ge \frac 1 8\dot{M}_{\rm{cr}} \\
	0, \qquad\qquad\quad |\dot{M}_{\rm{d}}|<\frac 1 8 \dot{M}_{\rm{cr}}
	\end{cases}
	\label{eq:3x}
\end{equation}
With the accumulation of helium on the surface of the WD, helium-shell flashes could occur if the mass of 
the helium layer reaches a critical value. Then the mass growth rate of the CO WD is 
\begin{equation}
	\dot{M}_{\rm{CO}}=\eta_{\rm{He}}\dot{M}_{\rm{He}},
	\label{eq:4}
\end{equation}
where $\eta_{\rm{He}}$ is the mass accumulation efficiency for helium-shell flashes. 
Its value depends on the mass of CO WD and $\dot{M}_{\rm{He}}$ (see more details in \citealt{kato2004}). 

\begin{figure}
	\centering
	\includegraphics[width=0.6\textwidth]{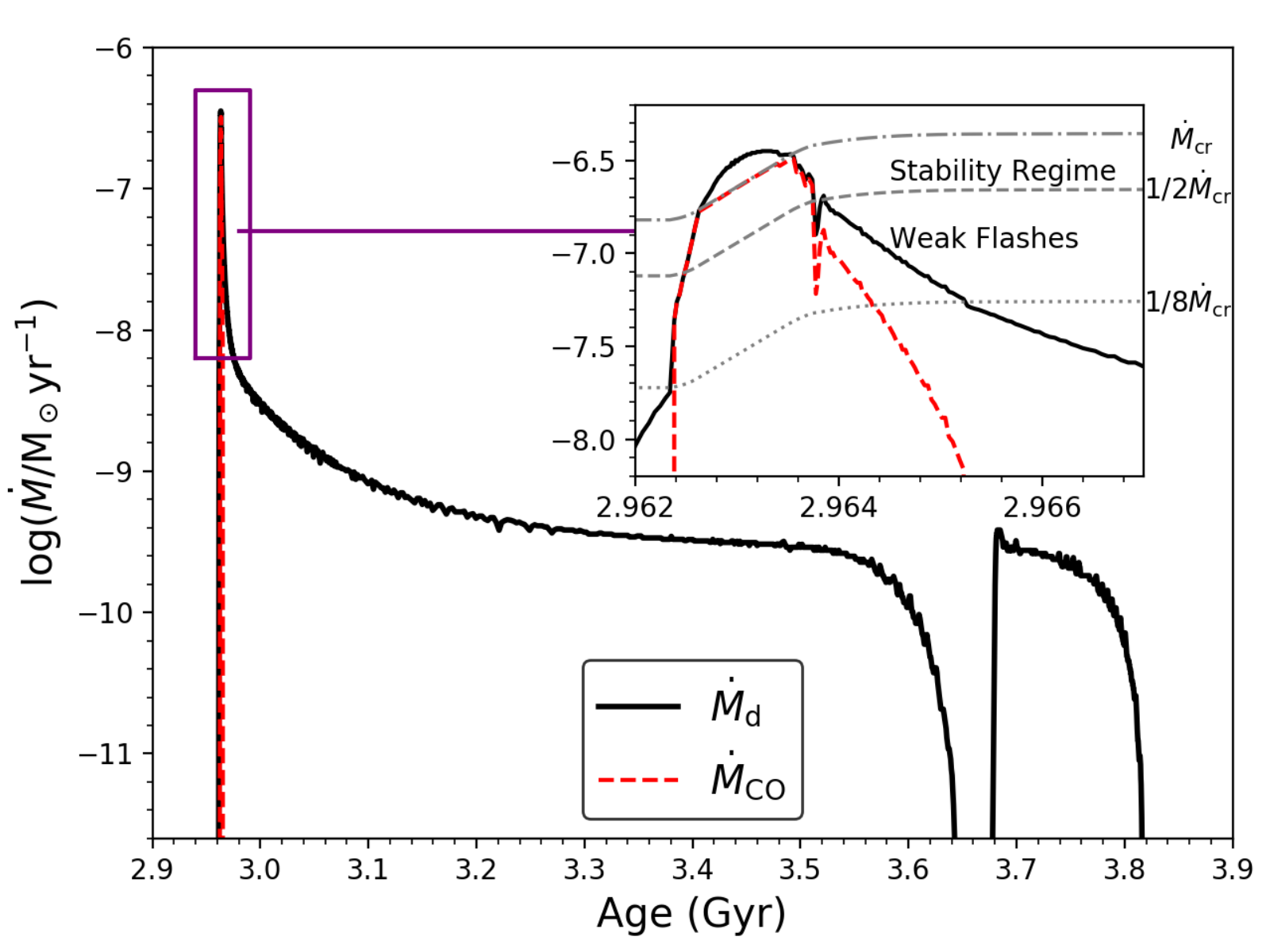}
    \caption{MT rate vs. star age for $M_{\rm{d,i}}=1.4 \;{\rm M_\odot}, 
	M_{\rm{CO,i}}=0.6\;{\rm M}_\odot,P_{\rm orb,i}=1.48\;\rm{d}$. The black solid line is the MT 
	rate, and the red dashed line shows the accumulation rate of the accretor. The gray dash-dotted 
	line in the inset is the critical MT rate $\dot{M}_{\rm cr}$, and the dashed and dotted lines 
	are $\frac 1 2\dot{M}_{\rm cr}$ and $\frac 1 8\dot{M}_{\rm cr}$, respectively. 
	See more details in the text.}
    \label{fig:3}
\end{figure}

Figure~\ref{fig:3} is an example to illustrate the CO WD accretion during MT process, 
where $M_{\rm d,i}=1.4\;{\rm M}_\odot, M_{\rm CO,i}=0.6\;{\rm M}_{\odot},P_{\rm orb,i}=1.48 \;\rm d$. 
The black solid and red dashed lines are for the MT rate, $\dot{M}_{\rm d}$, and the 
accumulation rate of CO WD, $\dot{M}_{\rm CO}$, respectively. Initially, the MT is on a thermal 
timescale, during which $\dot{M}_{\rm d}$ increases rapidly and exceeds $\frac 1 8\dot{M}_{\rm cr}$, 
$\frac 1 2\dot{M}_{\rm cr}$, $\dot{M}_{\rm cr}$, then deceases gradually after a while. The H-rich matter 
is accumulated at a rate of $\dot{M}_{\rm d}$ but limited by $\dot{M}_{\rm cr}$ in the areas of 
weak flashes, the stability regime, and the optically thick wind regime (above the dash-dotted line). 
In the stable regime, there is a difference between the red dashed and black solid lines during the decrease of 
$\dot{M}_{\rm d}$, since the mass increase of the CO WD is further constrained by $\eta_{\rm He}$, as shown 
in Equation~(\ref{eq:4}). When $\dot {M}_{\rm d}$ is below $\frac 1 8\dot{M}_{\rm cr}$, no mass can be accumulated 
onto the CO WD. The interruption of the MT process at the age of 3.65 Gyr is due to the discontinuity of 
the composition gradient during the first dredge-up stage (see also Section~\ref{subsubsec:5.1.3} and \citealt{jia2014}).

\subsubsection{Angular momentum loss}
\label{subsec:4.1.3}
We considered three physical processes of angular momentum loss in a binary, i.e. magnetic 
braking, GWR and ML. 
We use the formula derived by \citet{rappaport1983} to calculate the angular momentum loss by magnetic braking, 
\begin{equation}
	\dot{J}_{\rm{MB}}=-5.83\times 10^{-16}\frac{M_{\rm{env}}}{M_{\rm{d}}}\left(\frac{R_{\rm{d}}\omega_{\rm{spin}}}
	{\rm R_\odot \rm{yr}^{-1}}\right)^{\gamma_{\rm{MB}}}\rm M_\odot \rm R_\odot^2 \rm{yr}^{-2},
	\label{eq:5}
\end{equation}
where $\gamma_{\rm{MB}}=3$ in our simulation, $M_{\rm env}$ is the envelope mass of donor, $R_{\rm d}$ is the radius 
of donor, and $\omega_{\rm{spin}}$ is the spin angular velocity, which is equal to the orbital angular velocity 
$\omega_{\rm{orb}}$ as tidal synchronization is assumed. The magnetic braking effect is reduced if the convective envelope becomes 
too thin. So we turned on magnetic braking when the convective envelope fraction was 
larger than 0.01. Otherwise, magnetic braking was switched off.

The GWR plays a crucial role when orbital periods are less than several hr. The angular momentum loss 
due to GWR is \citep{landau} 
\begin{equation}
	\dot{J}_{\rm{GW}}=-\frac{32}{5c^2}\left(\frac{2\pi G}{P_{\rm{orb}}}\right)^{7/3}\frac{(M_{\rm{d}}M_{\rm{CO}})^2}
	{(M_{\rm{d}}+M_{\rm{CO}})^{2/3}},
	\label{eq:6}
\end{equation}
where $G$ is the gravitational constant, and $c$ is the velocity of light. 

The mass is assumed to be lost from the surface of CO WDs and take away the specific angular momentum of 
CO WDs. The angular momentum loss due to ML is 
\begin{equation}
	\dot{J}_{\rm{ML}}=-(1-\eta)\dot{M}_{\rm{d}}\left(\frac{M_{\rm{d}}}{M_{\rm{CO}}+M_{\rm{d}}}\right)^2 \frac{2\pi a^2}{P_{\rm{orb}}},
	\label{eq:7}
\end{equation}
where $a$ is the binary separation, and $\eta$ represents the total accumulation efficiency 
($\eta\equiv \eta_{\rm He}\eta_{\rm H}$). 
In our simulations, we did not consider the effect of spin-orbit coupling and tidal 
dissipation, and the orbit is assumed to be circular. 

\subsection{binary population synthesis}
\label{subsec:4.2}
There are four steps to performing the binary population synthesis. 
\begin{itemize}
\setlength{\itemsep}{0pt}
\setlength{\parsep}{0pt}
\setlength{\parskip}{0pt}
	\item[(1)] We generate $5\times 10^6$ primordial 
binaries by Monte Carlo simulation, evolve these binaries using the rapid binary-star 
evolution code \texttt{BSE} \citep{hurley00,hurley02}, and get a sample of binaries 
consisting of CO WD + MS/HG/RGB stars, in which the MS/HG/RGB stars just 
fill their Roche lobes and start transferring mass to the CO WD. From this step, we have 
four parameters for each system ($M_{\rm d},M_{\rm CO},P_{\rm RLOF},$ and $t_{\rm RLOF}$), where 
$P_{\rm RLOF}$ is the orbital period at the onset of RLOF, and $t_{\rm RLOF}$ is the 
age at the onset of RLOF. 
	\item[(2)] We interpolate these parameters in our 
grid from binary evolution calculation to get the physical quantities 
(e.g. $M_{\rm He},M_{\rm CO},P_{\rm orb},T_{\rm eff},\log g$) of ELM WDs in DDs from the 
RL channel. 
	\item[(3)] For the CE channel, of which the parameters of products are outside the parameter grid, 
the evolution of proto-He WD depends on the mass of H-rich 
layer above the He core, which is determined by the detailed CE ejection process (the most uncertain phase 
in binary evolution). In our study, we simply assume that the proto-He WDs from the CE channel have a 
similar structure to that from the RL channel. Based on the detailed binary evolution results (see 
Section~\ref{subsec:5.1}), we build a grid of models for the 
evolution of He WDs with a mass of 0.14-0.4 $\rm M_\odot$ from donor $M_{\rm d}=1.2\;\rm M_\odot$\footnote{According 
to Figure~\ref{fig:7} (Section~\ref{subsubsec:5.1.3}) shown below, for donor mass less than $1.5\;\rm M_\odot$, 
the envelope mass changes little with He WD mass, which indicates that the choice of donor mass has 
little effect on our results.}, 
in steps of $0.005\;\rm M_\odot$, starting from the end 
of MT. According to the core mass at the onset of the CE, we interpolate from the grid and obtain the following 
evolution of the products. 
	\item[(4)] We assume a constant star formation rate of 2 $\rm M_\odot~ yr^{-1}$ over 13.7 Gyr for the Galaxy 
\citep{chomiuk2011}, and combine the results of RL channel and CE channels to get the populations of ELM WDs in DDs. 
\end{itemize}

\subsubsection{Initial distribution for binary parameters}
\label{subsubsec:4.2.1}
The initial parameters for the Monte Carlo simulation are described as follows. The primary mass is given by the following 
initial mass function \citep{Miller1979, eggleton1989}:
\begin{equation}
	M = \frac{0.19X}{(1-X)^{0.75}+0.032(1-X)^{0.25}},
	\label{eq:8}
\end{equation}
where $X$ is a random number between 0 and 1, which gives the mass ranging from 
0.1 to 100 $\rm M_\odot$. This expression has gained support from the follow-up study \citep{kroupa1993}. 
The initial mass ratio distribution is taken as a constant distribution \citep{mazeh1992}, i.e. $n(q')=1$, 
$0\leq q' \leq 1$, where $q'$ represents the mass ratio of initial binary. 
The distribution of initial separation is a uniform distribution in 
$\log a$ for wide systems and a power-law distribution at close separation \citep{han1998}:
\begin{equation}
	an(a)=
	\begin{cases}
	0.07(a/a_0)^{1.2},\qquad a\le a_0 \\
	0.07, \qquad \qquad \quad a_0 \le a \le a_1,
	\end{cases}
	\label{eq:9}
\end{equation}
where $a_0=10\;\rm{R}_{\odot}$, $a_1=5.75\times 10^6\;\rm{R}_{\odot}$. This distribution gives approximately 50 
percent of systems with orbital periods less than 100 yrs. 

\subsubsection{The CE channel}
\label{subsubsec:4.2.2}
For ELM WDs from the CE channel, we simply assume that the binary with a CO WD enters CE phase if the MT 
rate is larger than $10^{-4}\;\rm M_\odot yr^{-1}$ (see Section~\ref{subsubsec:4.1.1}), 
and adopt standard energy budget formula for the CE phase 
\citep{webbink1984,livio1988,dekool1990}, that is
\begin{equation}
    \alpha_{\rm{CE}}\left(\frac{GM_{\rm{core}}M_{\rm{CO}}}{2a_{\rm{f}}}-\
	\frac{G(M_{\rm{core}}+M_{\rm env})M_{\rm{CO}}}{2a_{\rm{i}}}\right)=\
	\frac{M_{\rm{d,i}}M_{\rm env}}{\lambda R_{\rm d,i}},
	\label{eq:10}
\end{equation}
where the left side is the release of orbital energy, and the right side is the bind energy of envelope, 
$M_{\rm core}$ and $M_{\rm env}$ are the core mass and the envelope mass of the donor,
$\alpha_{\rm CE}$ and $\lambda$ are the CE ejection efficiency\footnote{The contribution of internal energy 
is small when the giant is near the base of giant branch \citep{han1994,chen2017}, and is not considered here. 
} and the envelope structure parameter, respectively. 
We simply set $\lambda=1$, $\alpha_{\rm CE}=0.25$, 0.5, or 1.0. The model of $\alpha_{\rm CE}=1$ is 
referred to be the standard model in this paper. 

\subsubsection{Distinguishing the sample from different evolutionary channels}
\label{subsubsec:4.2.3}
In order to compare our results with those of observations, we need to divide the clean sample into 
two groups according to their evolutionary channels. Here we introduce a method to distinguish the sample 
as below. The basic idea is that all the ELM WDs in DDs are assumed to originate from the CE channel, which 
results in a CE coefficient, $\alpha_0$, for each sample, and those with unreasonable values of $\alpha_0$ 
are considered to be produced from the RL channel. 

To do this, we firstly evolve a $1.2\;\rm M_\odot$\footnote{
The reason to choose $M_{\rm d}=1.2\;\rm M_\odot$ is according to Figure~\ref{fig:13} (Section~\ref{subsubsec:5.2.3}), 
that the most common progenitors for ELM WDs produced from the CE channel have masses in the range of 
$0.95-1.25\;\rm M_\odot$.} star from MS to RGB until its core mass reaches $\sim0.4\;\rm M_\odot$. We 
simply assume that the ELM WD mass from the CE channel is equal to the core mass at the onset 
of MT process and that the companion has not accreted any material during the CE process. For 
each observed ELM WD in DD, we then obtain the stellar radius, $R_{\rm d}$, the envelope mass, $M_{\rm env}$, 
and the binding energy\footnote{The bind energy $E_{\rm{bind}}$ 
is calculated by considering the full stellar structure \citep{han1994,dewi2000}, and we define the core boundary 
at the position where hydrogen mass fraction equals to 0.1.} at a given core mass $M_{\rm core}$ ($=M_{\rm He}$) 
along the evolutionary track. We further have the initial separation \citep{eggleton1983} 
\begin{equation}
	a_{\rm{i}}=R_{\rm RL}\frac{0.6q^{2/3}+\ln (1+q^{1/3})}{0.49q^{2/3}},
	\label{eq:12}
\end{equation}
where $R_{\rm RL}=R_{\rm d}$, $q={M_{\rm d}}/M_{\rm a}$ ($M_{\rm d}=1.2\;\rm M_\odot$, $M_{\rm a}$ is 
the companion mass and is given by observations). Finally, the value of $\alpha_0$ is calculated by 
\begin{equation}
	\alpha_{\rm{0}}=\frac{E_{\rm{bind}}}{G(M_{\rm{core}}+M_{\rm{env}})M_{\rm{a}}/(2a_{\rm{i}})-\
	GM_{\rm{core}}M_{\rm{a}}/(2a_{\rm{f}})},
	\label{eq:11}
\end{equation}
where $a_{\rm f}$ is the final separation\footnote{
In fact, after the ending of the CE channel, the orbital periods will be decreased due to the angular momentum being 
taken away by magnetic braking and GWR. Therefore, the post-CE periods are larger than the observed periods, then the 
true value of CE efficiency is larger than $\alpha_{\rm 0}$ calculated above \citep{zorotovic2010}. However, 
it has little effect on our discussion, so we neglect this effect in this work.} and is given by observation. 

\begin{figure}
	\centering
	\includegraphics[width=0.6\textwidth]{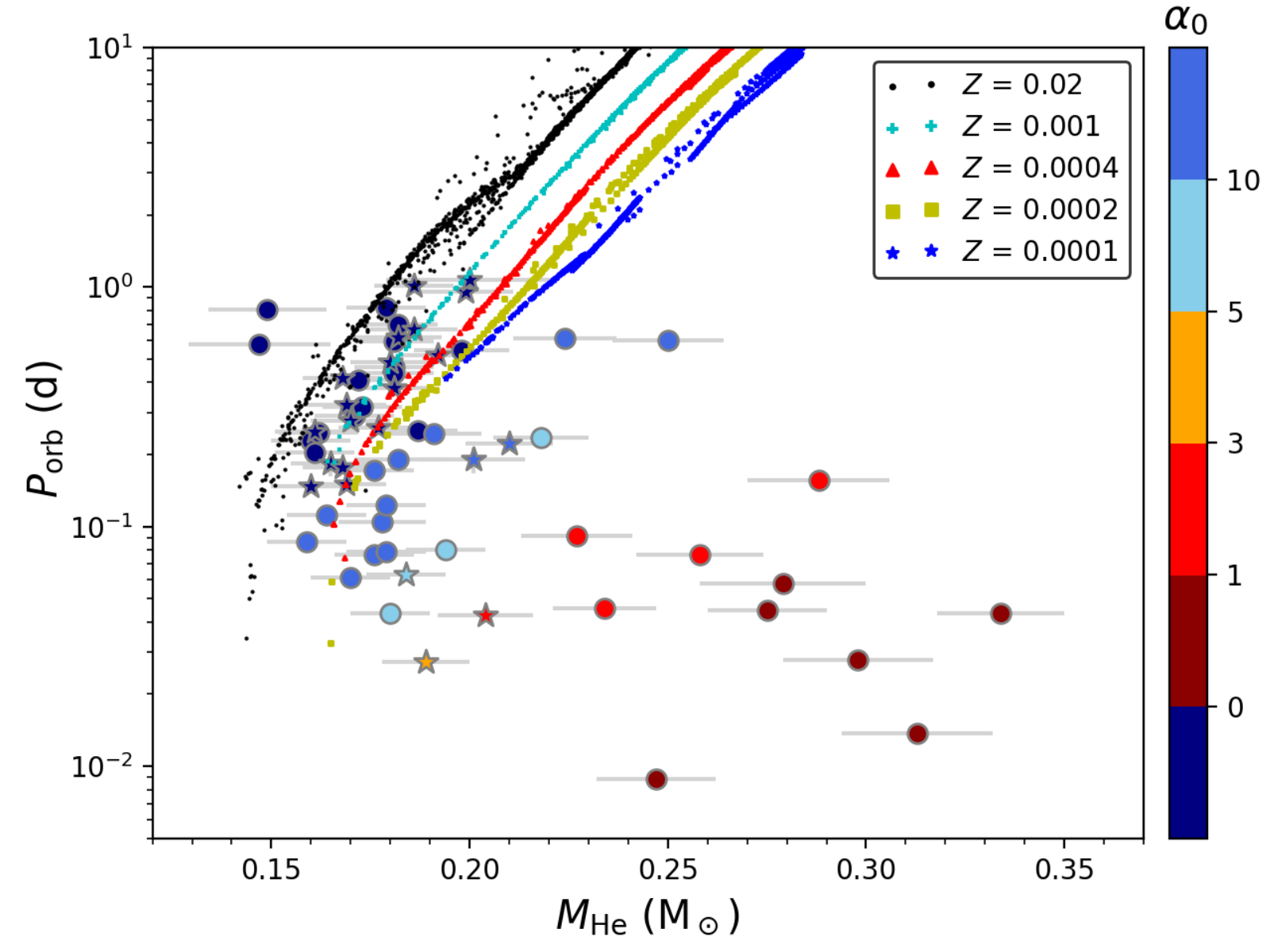}
	\caption{Distinguishing the formation channel for the observed samples by $\alpha_{\rm 0}$. 
	The values of $\alpha_{\rm 0}$ are obtained by assuming that all of the observed systems are 
	produced from the CE channel; see the text for the detailed calculation methods. 
	The circles and stars with error bars are for the disk and halo objects in ELM Survey, respectively, 
	where the colors indicate the value of $\alpha_{\rm{0}}$. 
	The cyan plus, red triangles, yellow squares, and blue stars are from our simulations for 
	metallicities $Z=0.001$, $0.0004$, $0.0002$, and $0.0001$, respectively. 
	We set the critical value between the CE and RL channels to be simply $\alpha_{\rm 0}=5$. 
	Therefore, the colors (navy, royal, sky blue) with $\alpha_{\rm{0}}<0$ and $\alpha_{\rm 0}>5$ 
	are for systems produced from RL channel, and other colors (orange, red, and dark red) with $0<\alpha_0<5$ 
	are for those formed from CE channel. 
	} 
    \label{fig:4}
\end{figure}

In Figure~\ref{fig:4} we present the value of $\alpha_{\rm{0}}$ for different samples in the 
$M_{\rm{He}}-P_{\rm orb}$ plane. Filled circles with 
error bars are for the observed systems. Navy blue marks $\alpha_{\rm{0}}<0$, which means that 
the final orbit separation after CE phase is longer than the initial separation. 
Royal blue and sky blue are for the cases of $\alpha_{\rm{0}}>10$ and $10>\alpha_{\rm 0}>5$, 
which means that these systems are also difficult to form through CE channel. In other words, 
the assumption that these systems formed via CE channel is unreasonable. 
Other colors are for $\alpha_{\rm 0}$ from 0 to 5, which are more likely formed from the CE channel. 
Systems with lower metallicity ($Z=0.001,0.0004,0.0002,0.0001$) 
are also presented, and it is clear that the orbital period is positively correlated with the metallicity 
\citep{nelson2004}. The low-metallicity systems match the observations better, 
we will discuss the effect of metallicity in Section~\ref{subsubsec:5.2.1}. 
In this work, we set the critical value between CE and RL channels as $\alpha_{\rm 0}=5$ artificially. 

\section{Results}
\label{sec:5}
\subsection{Results of Detailed Binary Evolution}
\label{subsec:5.1}
\subsubsection{Parameter space for producing ELM WDs from the RL channel}
\label{subsubsec:5.1.1}
\begin{figure}
	%\subfigure{
	%\includegraphics[width=\columnwidth]{fig_4}}
	%\subfigure{
	%\includegraphics[width=\columnwidth]{fig_4_2}}
	\centering
	\includegraphics[width=0.5\textwidth]{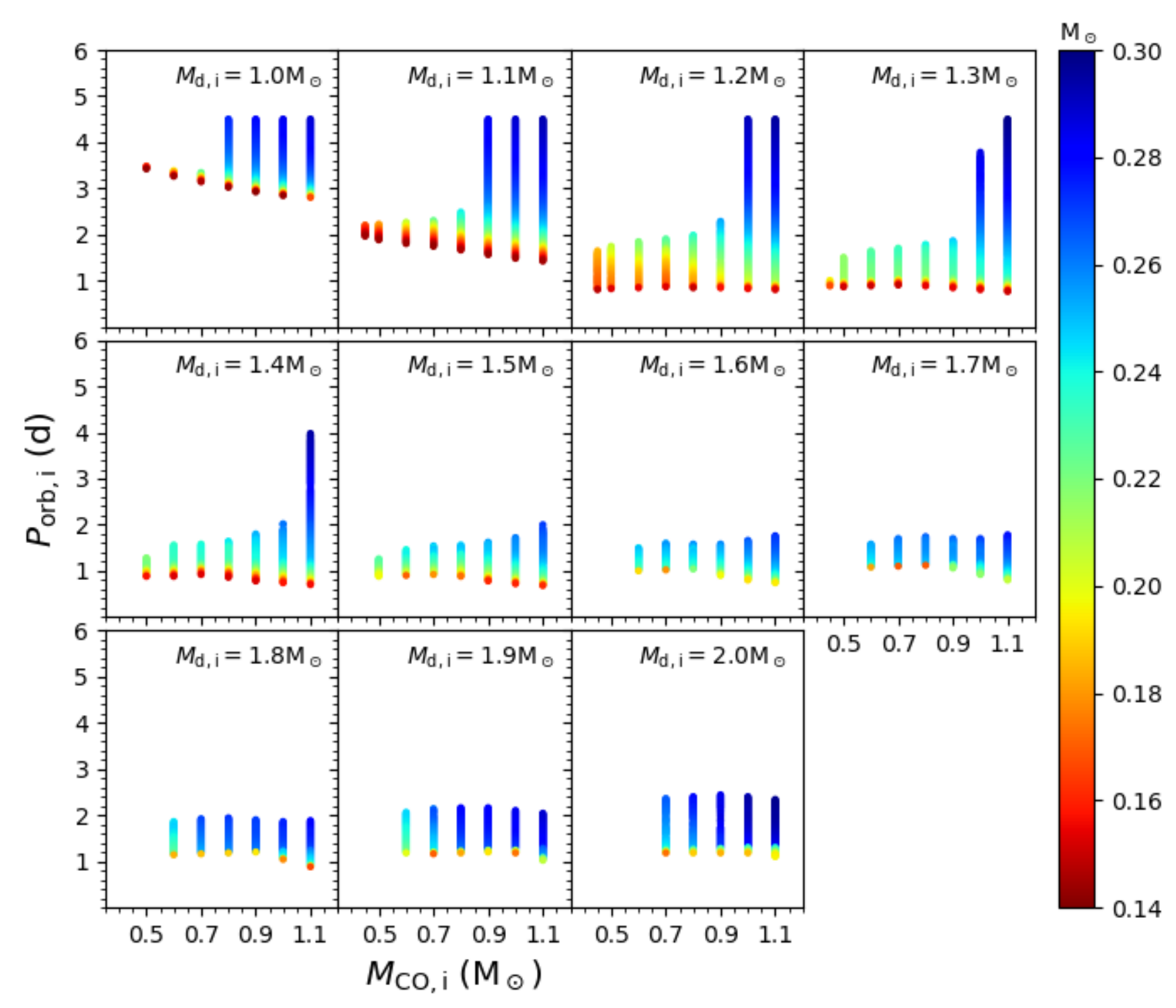}
	\vfill
	\includegraphics[width=0.5\textwidth]{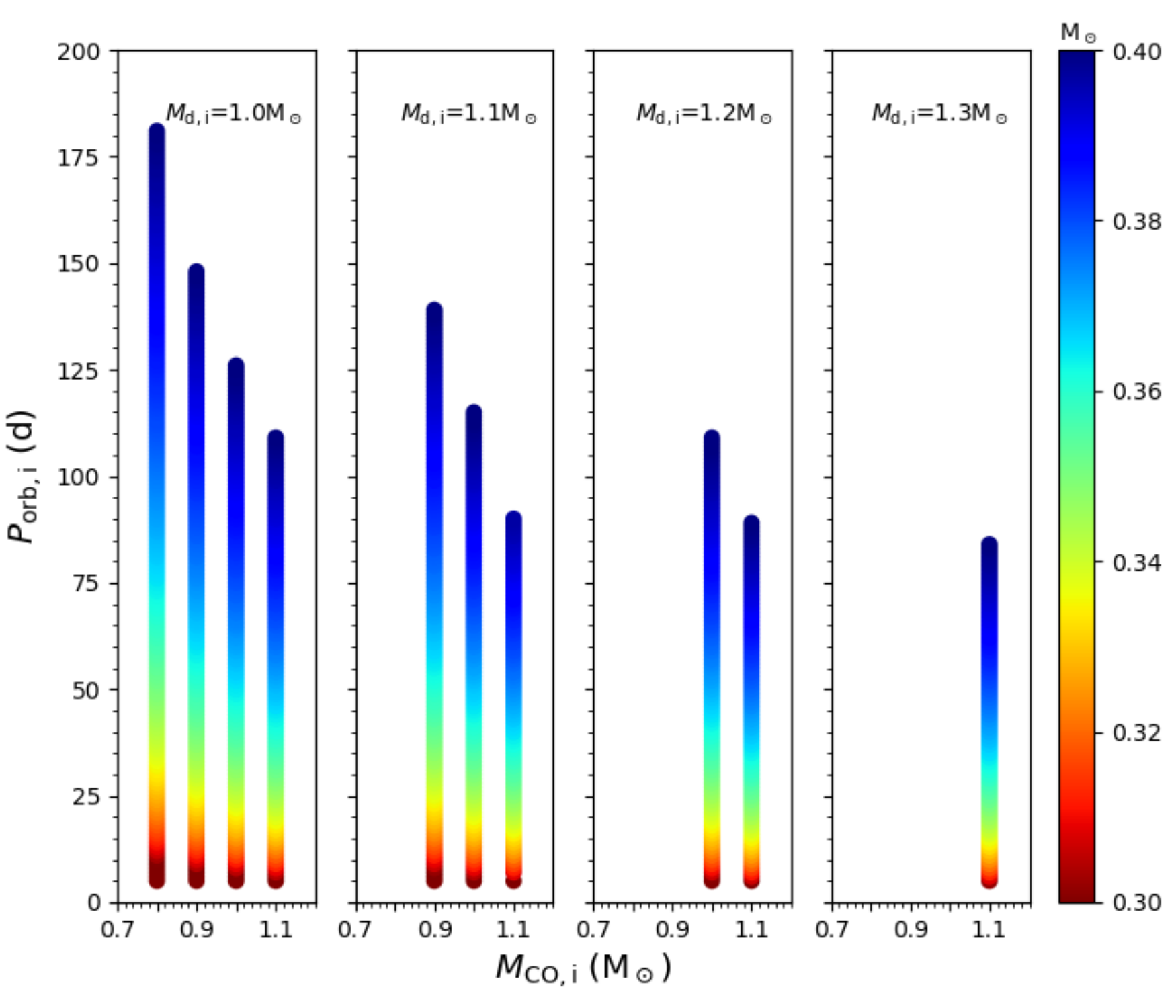}
	\caption{
	The parameter space for producing ELM WDs from the RL channel in the CO WD mass - initial 
	orbital period plane. The initial donor mass is indicated in each panel. For clarity, the 
	upper panel only shows the case for initial orbital period less than 4.5 d, where the lower boundary 
	is determined by $P_{\rm b}$ and the upper boundary is determined by the maximum MT rate 
	of $10^{-4}\;\rm M_\odot yr^{-1}$. When the donor mass is lower than 1.4 $\rm M_\odot$, the He WD may also 
	be formed via the RL channel when $P_{\rm orb,i}>4.5\;\rm d$ and $M_{\rm CO,i}>0.7\rm \;M_\odot$. 
	The parameter space for this part is shown in the bottom panel, where 
	the upper boundary is determined by the maximum He WDs of $0.4\;\rm M_\odot$. 
	Note that the color scale of two panels is different. 
	}
	\label{fig:5}
\end{figure}

Based on the calculation method introduced above, we get the parameter space for producing ELM WDs from 
the RL channel. Figure~\ref{fig:5} shows the CO WD mass - initial orbital period plane, where the colors indicate 
the final He WD mass, and the initial donor mass is indicated in each panel. 
For clarity and comparison, we only show the case for $P_{\rm orb,i}<4.5\;\rm d$ in the upper panel, where the 
lower boundary is determined by bifurcation period, $P_{\rm b}$, 
and the upper boundary is determined by the maximum MT rate of $10^{-4}\;\rm M_\odot yr^{-1}$. When the donor 
mass is lower than 1.4 $\rm M_\odot$, the He WD may also be formed via the RL channel when $P_{\rm orb,i}>4.5\;\rm d$ 
and $M_{\rm CO,i}>0.7\;\rm M_\odot$. The parameter space for this part is shown in the bottom panel, where 
the upper boundary is determined by the maximum He WDs of $0.4\;\rm M_\odot$. 
We can see that the bifurcation period for $M_{\rm{d,i}}=1.0,1.1\;\rm M_\odot$ is obviously 
larger than that for $M_{\rm{d,i}}\ge 1.2\;\rm M_\odot$, which is caused by magnetic braking. 
For $M_{\rm{d,i}}\lesssim 1.15\;\rm M_\odot$, the donors have a convective envelope on the MS, and magnetic braking 
takes away the orbital angular momentum and makes the orbit shrink in advance \citep{chen2017}. 
Besides, for $M_{\rm d,i}\geq1.6\;\rm M_\odot$, there are few ELM WDs with a mass less than 
$\sim$$0.18\;\rm M_\odot$, since the helium core of these donors grows too rapidly \citep{sunm2017}. 
It is noteworthy that the maximum initial period decreases with the increasing CO WD mass in the 
bottom panel, due to the fact that MT process at the onset of RLOF becomes moderate 
with the increase of CO WD mass; i.e., initial mass ratio (the accretor mass to donor mass) decreases, 
and the core in the donor has more chances to increase during the MT process, in comparison to 
the case of low-mass CO WDs.

\subsubsection{Typical evolutionary tracks}
\label{subsubsec:5.1.2}
\begin{figure}
	\centering
	\includegraphics[width=0.55\textwidth]{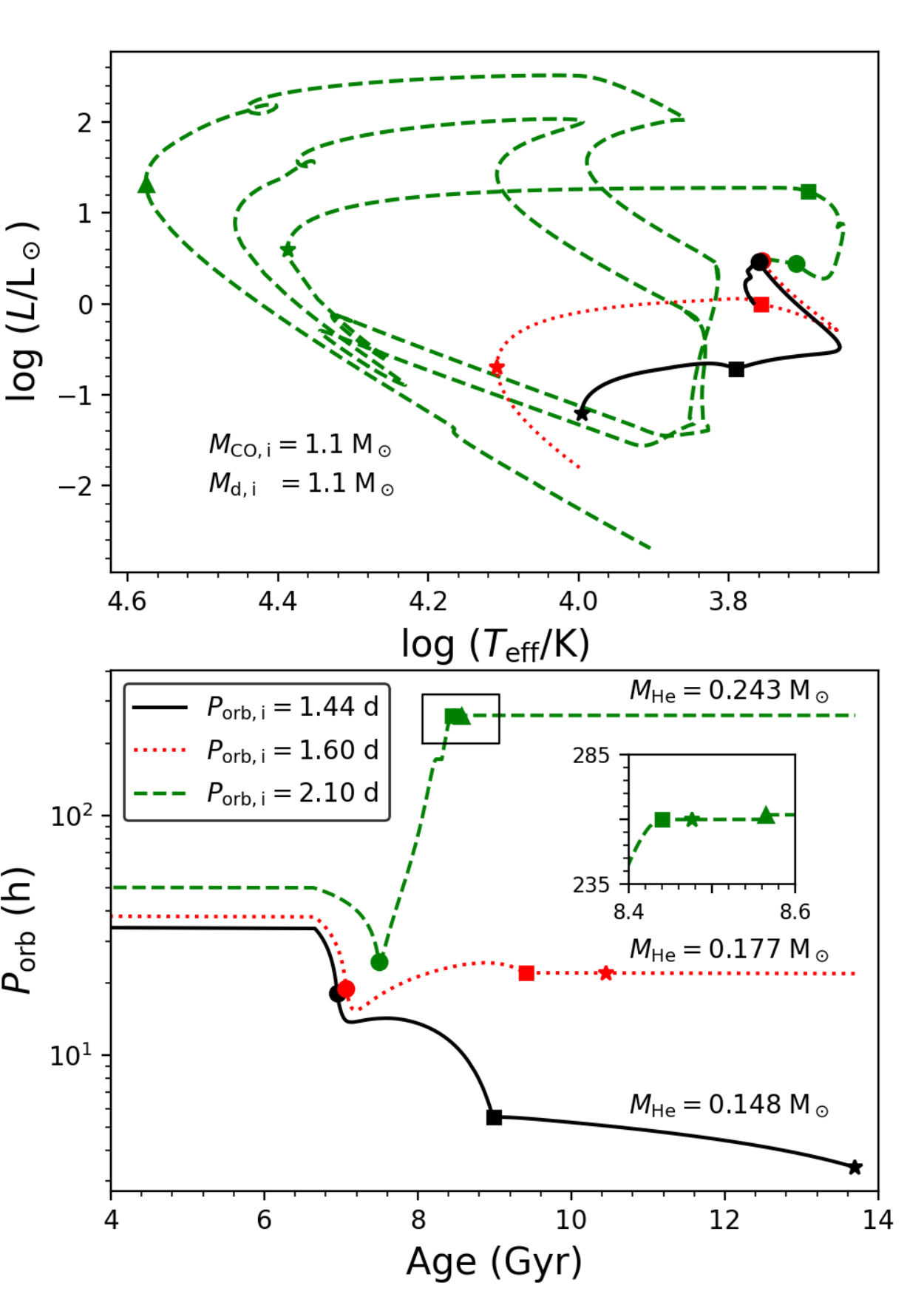}
    \caption{Three typical evolutionary tracks on the Hertzsprung-Russel diagram and orbital period 
	evolution for binaries with $M_{\rm{d,i}}=1.1\;\rm M_\odot$ and $M_{\rm{CO,i}}=1.1\;\rm M_\odot$. 
	The solid, dotted, and dashed lines correspond to initial period of 
	$P_{\rm orb,i} = 1.44,1.60,2.10\;\rm{d}$, respectively. The circles, squares, stars and triangles 
	are for the beginning and end of MT, the termination of nearly 	constant luminosity and the 
	maximum temperature before cooling, respectively. (Note: the maximum temperature is the temperature after the 
	loop of H-shell flashes)}
    \label{fig:6}
\end{figure}

We select three typical evolutionary tracks of binary systems with 
$M_{\rm d,i}=1.1\;\rm M_\odot, M_{\rm CO,i}=1.1\; \rm M_\odot$ from the parameter space as 
shown in Figure~\ref{fig:6}. 
The solid, dotted, and dashed lines are for initial orbital periods $P_{\rm orb,i}=1.44, 1.60 $ and $2.10\;\rm d$, 
respectively. For the first two cases, the donors start to transfer mass to the companions after leaving the 
MS, and they do not ascend the RGB. 
After the end of MT, proto-He WDs do not enter into the cooling stage immediately 
because the residual hydrogen layer is still burning, sustaining a relatively high 
luminosity \citep{webbink1975}. The donors become proto-He WDs 
and enter into a nearly constant luminosity phase. 
For $P_{\rm orb,i}=1.44\; \rm{d}$, the proto-He WD 
has reached the maximum age, 13.7 Gyr, without entering into cooling stage. 
For the latter case with a relatively larger initial orbital period, i.e. $P_{\rm orb,i}=2.10 \;\rm{d}$, the donor 
ascends the RGB during MT phase and eventually leaves a relatively massive core 
($0.243\;\rm M_\odot$). After the nearly constant luminosity phase, two strong H-shell flashes appear in the 
H-rich envelope, then the proto-He WD enters into the cooling phase. 

The symbols in the figure show some key points during the evolution, i.e. the onset of MT 
(circles), the end of MT (squares), the termination of nearly constant 
luminosity (stars), and the maximum temperature, $T_{\rm max}$, before cooling 
(triangles). Note that $T_{\rm max}$ is the temperature after the loop of H-shell flashes. 
We define the timescale of contraction phase $t_{\rm{c}}$ as the time spent 
between the end of MT and the maximum temperature before H-shell flashes\footnote{The 
definition of $t_{\rm c}$ in this paper is different from that of \citet{istrate2014b}, who 
included the timescale of H-shell flashes.}\citep{chen2017}. 

The lower panel shows the evolution of orbital period as a function of stellar age. 
We see that the period has decreased 
before MT occurs, since the magnetic braking plays a role in the last phase of MS. During the MT phase, 
the envelope mass decreases until the convective envelope disappears, then the magnetic braking 
becomes invalid and the donor contracts into the RL, leading to the detachment of binary. 
After that, the angular momentum loss is driven by GWR, which becomes important when 
$P_{\rm orb}\lesssim0.2\;\rm d$ as shown by the track of system with $P_{\rm orb,i}=1.44\;\rm d$. 
The inset shows the timescale of contraction phase and the H-shell flashes for systems with 
$P_{\rm orb,i}=2.10\;\rm d$, which are on the order of $10^{8}$ yr. 
The final He WD masses of the binary systems are 0.148, 0.177, 0.243$\;\rm M_\odot$, respectively, as 
indicated in the lower panel. It can be observed 
visually that there is a strong correlation between $M_{\rm{He}}$ and $t_{\rm{c}}$, where 
lower-mass proto-He WDs have a longer lifetime at the contraction stage (see Section~\ref{subsubsec:5.1.3}). 

\subsubsection{Dependence of contraction timescale on proto-He WD mass}
\label{subsubsec:5.1.3}
\begin{figure}
	\centering
	\includegraphics[width=0.55\textwidth]{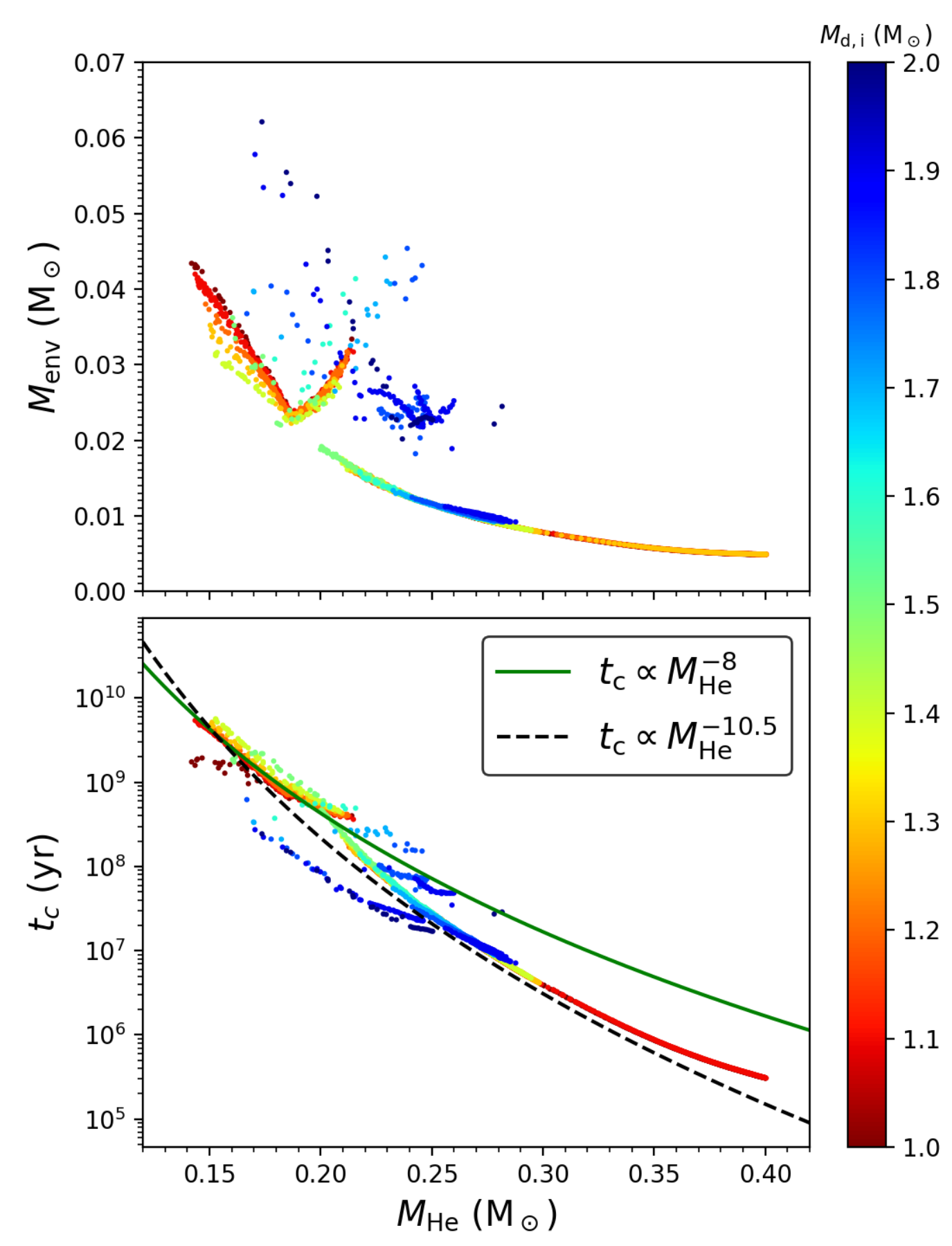}
    \caption{Dependence of envelope mass (upper panel) and contraction timescale (lower panel) 
	on proto-He WD mass. We define $M_{\rm{env}}$ as the envelope mass at the end of MT and 
	$t_{\rm c}$ is defined in Section~\ref{subsubsec:5.1.2}. 
	The solid line and dashed line correspond to the fitting formula given by \citet{chen2017}. 
	}
    \label{fig:7}
\end{figure}
From the above discussion, we see that there is a strong correlation between $M_{\rm He}$ and $t_{\rm c}$. 
To illustrate this phenomenon, we present the dependence of envelope 
mass, $M_{\rm env}$\footnote{The He core boundary defined in \texttt{MESA} is at the position where 
hydrogen mass fraction equals to 0.01.}, and contraction timescale, $t_{\rm c}$, on 
$M_{\rm{He}}$ in Figure~\ref{fig:7}. The colors indicates the initial donor masses. 
The results shown here are similar to those of EL CVn-type binaries and millisecond pulsar binaries 
\citep{chen2017,istrate2016b,istrate2016a}, i.e. the envelope mass (and the timescale for contraction) are 
(strongly) anticorrelated with the He WD mass, and there is a small upturn for $M_{\rm env}$ when 
$M_{\rm He}\gtrsim0.19\;\rm M_\odot$ due to stellar contraction at the discontinuity composition gradient induced by 
the first dredge-up (see discussion in \citealt{chen2017}). The strong anticorrelation between $t_{\rm c}$ 
and $M_{\rm He}$, i.e. proportional to $M_{\rm He}^{-8}$ or $M_{\rm He}^{-10.5}$, suggests that low-mass He 
WDs are more likely to be observed (see Section~\ref{subsec:5.2} for more). The 
dispersion in the figure results from the products of $M_{\rm d,i}\gtrsim 1.6\;\rm M_\odot$, which have 
nondegenerate cores and the termination of MT is a gradual process rather than a rapid contraction, as 
that in low-mass donors. 

\subsubsection{Comparison of the model grid with observations}
\label{subsubsec:5.1.4}
The observed properties of ELM WDs could be used to examine the reliability of our binary evolution 
calculation results. Here we compare our results to the observations of $T_{\rm eff}-\log g$ plane 
and $M_{\rm He}-M_{\rm CO,f}$ planes. 

\begin{figure}
	\centering
	\includegraphics[width=0.6\textwidth]{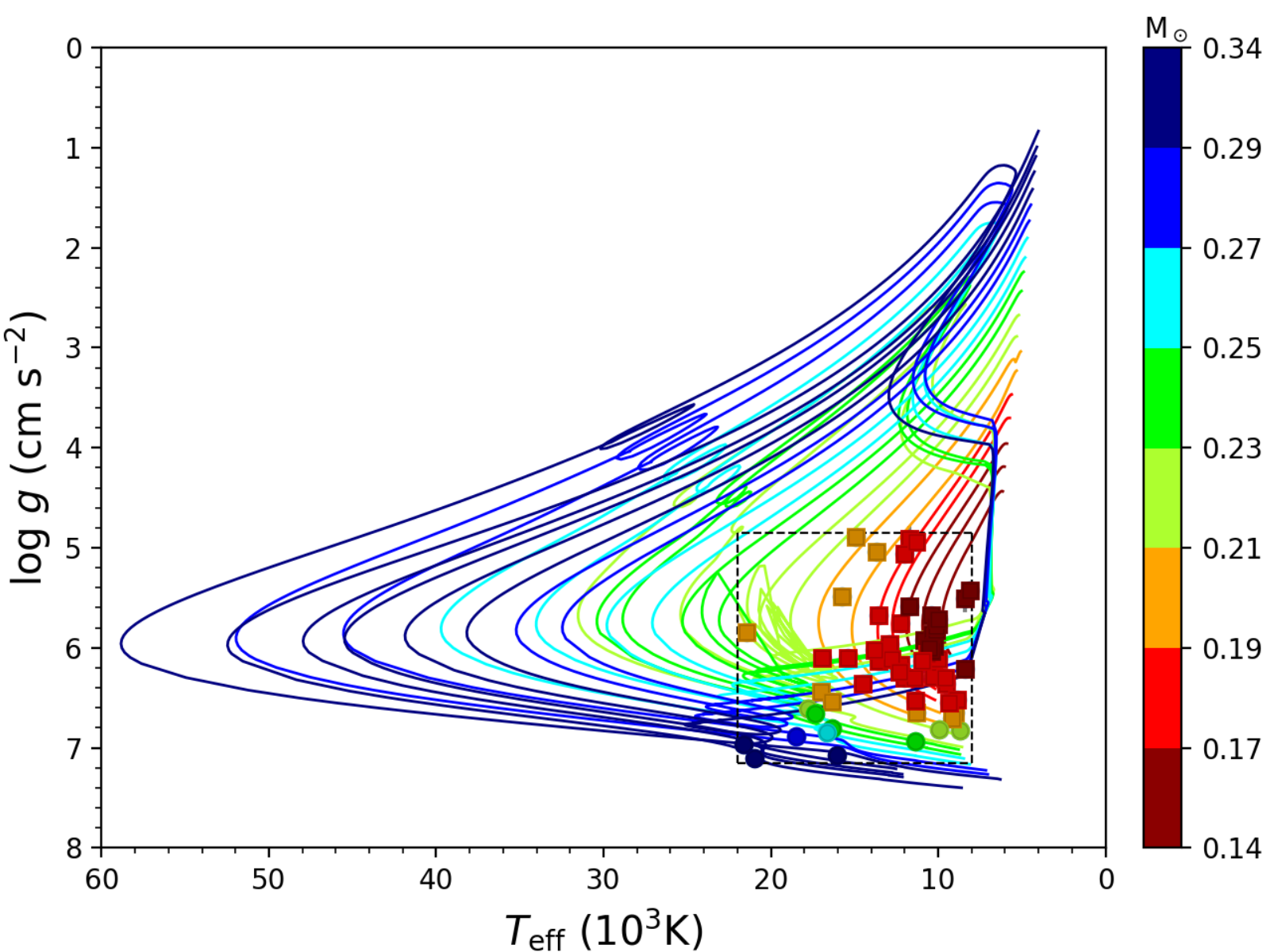}
    \caption{Some evolutionary tracks in the $T_{\rm{eff}}-\log g$ plane. 
	The colors represent the mass of He WDs, and 
	the observed He WDs with $M_{\rm He}\lesssim 0.21 \;\rm M_\odot$ and 
	$M_{\rm He}\gtrsim0.21\;\rm M_\odot$ are denoted by filled squares and circles, 
	respectively. 
	The vertical and the horizontal dashed lines give the ranges for effective temperature 
	and surface gravity from the ELM Survey, i.e. $4.85<\log g<7.15$ and $8000<T_{\rm eff}<22000\;\rm K$. 
	}
    \label{fig:8}
\end{figure}

Figure~\ref{fig:8} shows the selected evolutionary tracks in the $T_{\rm{eff}}-\log g$ plane 
with $M_{\rm He}$ from $0.145$ to $0.335\; \rm M_\odot$ in steps of $0.01\;\rm M_\odot$ 
in our calculation. These tracks show the evolution of He WDs from the termination of MT 
to the maximum age. The vertical and the horizontal dashed lines give the ranges for effective temperature 
and surface gravity from the ELM Survey, i.e. $4.85<\log g<7.15$ and $8000<T_{\rm eff}<22000\;\rm K$. 
As shown in the figure, most of the samples with $M_{\rm He}\lesssim 0.21\rm M_\odot$ 
(squares) are in the contract phase and are bloated somehow, i.e., with lower 
surface gravity and the relatively long timescale in this phase predicted by 
theoretical studies \citep{istrate2014b,chen2017}. Meanwhile, all the samples 
with $M_{\rm He}\gtrsim 0.23\;\rm M_\odot$ (filled circles) are located 
below the turnoff at the large temperature end due to the much longer timescale in 
this phase for He WDs with such masses in comparison to that in the nearly constant 
luminosity phase as shown in Figure~\ref{fig:6} (see also \citealt{istrate2014b}). 
We noticed that the samples in the contraction phase are close to the turnoff of the 
maximum temperature rather than homogeneously distributed in this stage, which could also be 
understood by evolutionary timescale. From Figure~7 of \citet{chen2017} we see 
that the proto-He WDs evolve significantly faster during the constant luminosity phase in comparison to 
that around the turnoff. 

\begin{figure}
	\centering
	\includegraphics[width=0.6\textwidth]{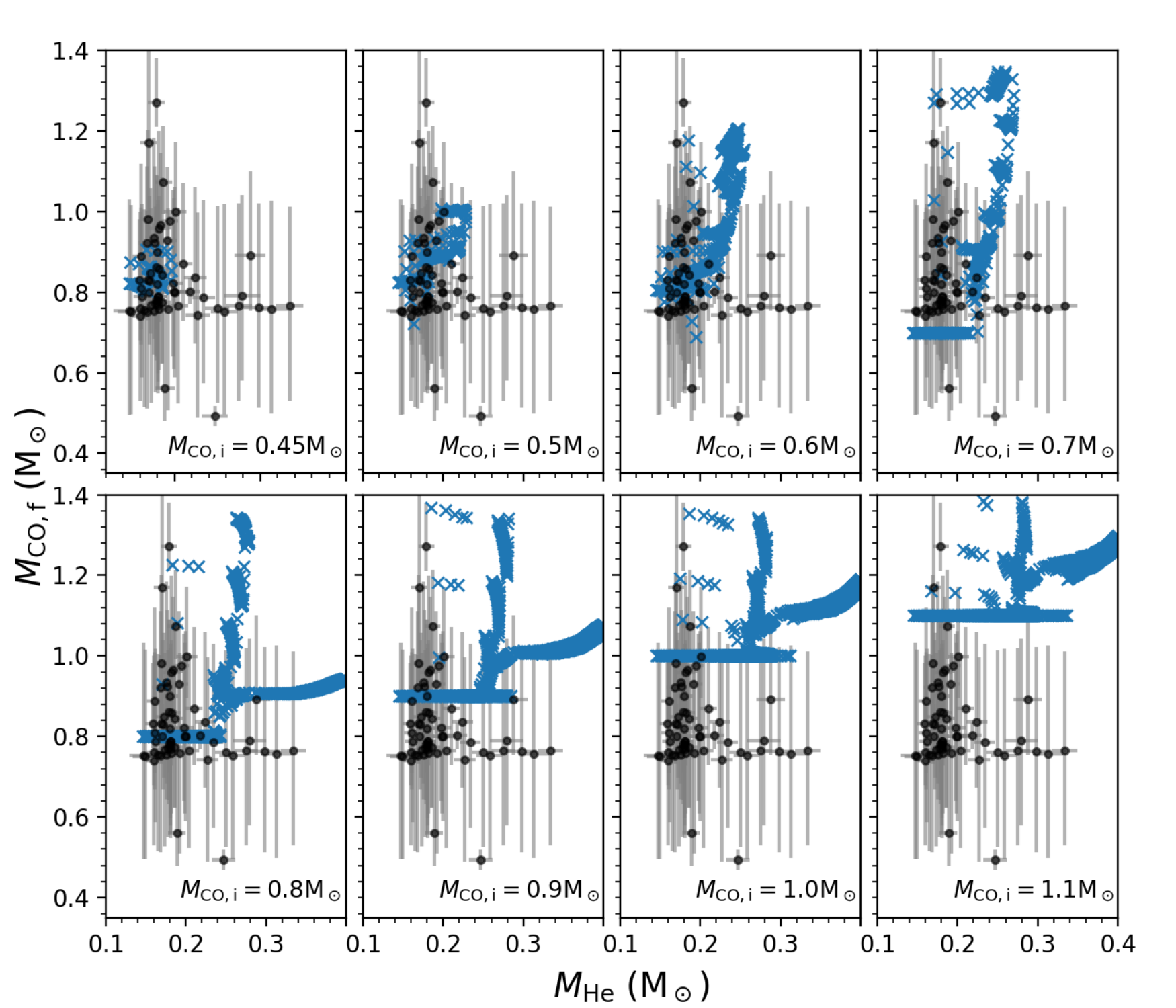}
    \caption{The $M_{\rm{He}}-M_{\rm{CO,f}}$ plane in our calculations. The observed samples and our grid 
	results are shown with black circles and blue crosses, respectively. The initial CO WD masses are 
	indicated in each panel. 
	}
	\label{fig:9}
\end{figure}
To understand the mass growth of the accretors, we plot the $M_{\rm{He}}-M_{\rm{CO,f}}$ diagram 
in Figure~\ref{fig:9}, where $M_{\rm CO,f}$ denotes the final mass of the accretors, and the 
observed samples are also shown. The initial CO WD mass is indicated in each panel. 
Most low-mass accretors ($M_{\rm CO,i}\lesssim 0.6\;\rm M_\odot$) could increase in mass about 
$0.2-0.6\;\rm M_\odot$ and have a final mass in the 
range of $0.8-1.2\rm M_\odot$. However, the initially massive CO WDs ($\geq 0.7\rm\;M_\odot$) hardly 
grow in mass when $M_{\rm He}$ is less than a certain value as shown in the figure. This can be 
understood as follows. The critical MT rate is larger for massive CO WDs according to Equation~(\ref{eq:2}), and 
the systems with short $P_{\rm orb,i}$ (or low $M_{\rm He}$) start MT earlier. The MT process 
is moderate, i.e. the MT rate $\dot{M}_{\rm d}$ is low in comparison to those with long $P_{\rm orb,i}$ 
for a given $M_{\rm d}$ and $M_{\rm CO,i}$. So, when $P_{\rm orb,i}$ (or $M_{\rm He}$) is less than 
some certain value, the CO WDs do not accrete any material. With the increasing of $P_{\rm orb,i}$, 
$\dot{M}_{\rm d}$ increases, and the CO WDs could accrete some material and grow in mass. 

\subsection{Binary population synthesis results}
\label{subsec:5.2}
According to the assumption in Section~\ref{subsec:4.2}, we get the statistical properties of ELM WDs in DDs 
for the Galaxy, including the He WDs, CO WDs and progenitors mass distribution, and the birth rate and 
local space density of these systems. It is noted that our simulations have considered the evolutionary 
timescale of ELM WDs, as well as the GWR merger timescale of binary systems, 
so the results should be directly used to compare with the observations. 

\subsubsection{Mass distribution of ELM WDs}
\label{subsubsec:5.2.1}
\begin{figure}
	\centering
	\includegraphics[width=0.5\textwidth]{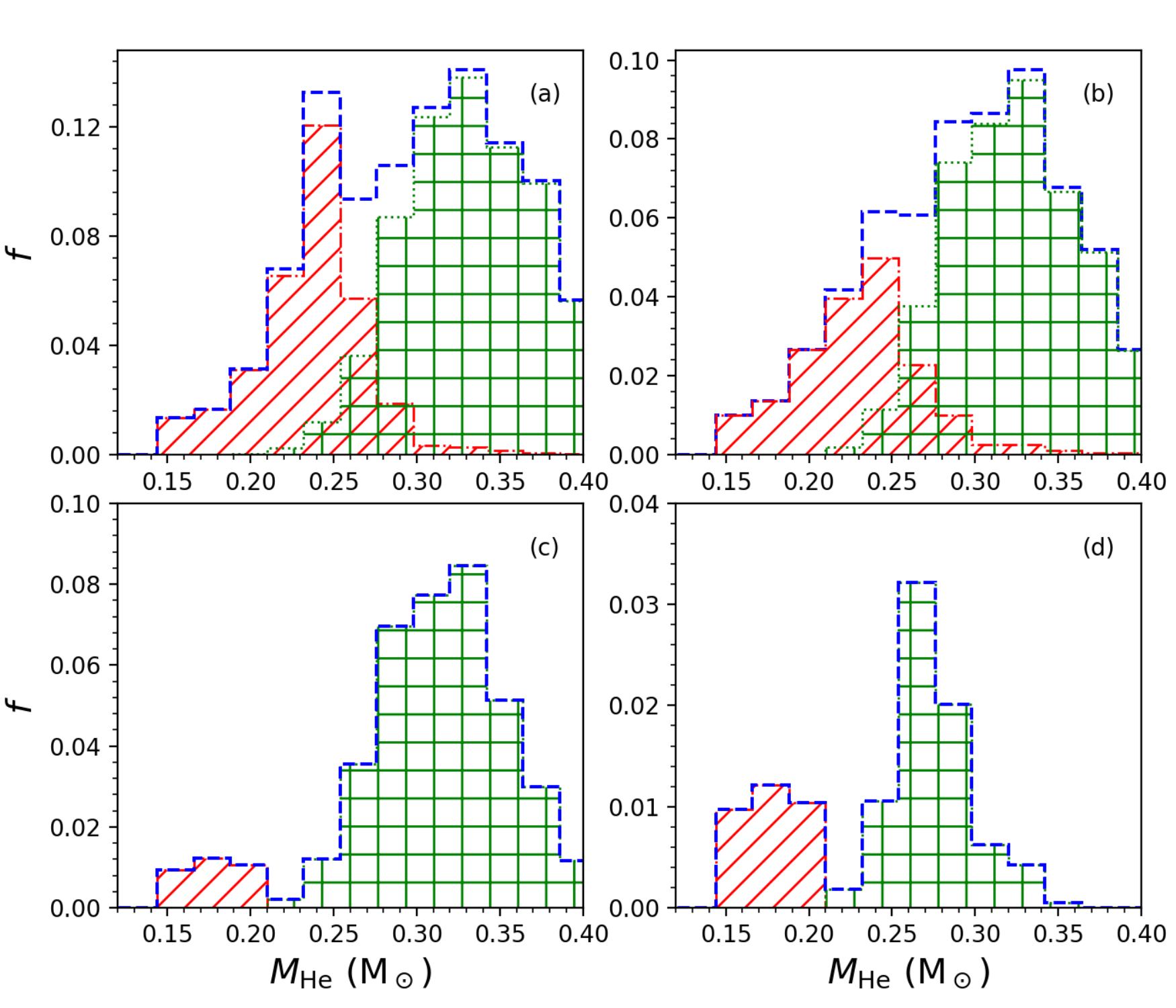}
	\vfill
	\includegraphics[width=0.5\textwidth]{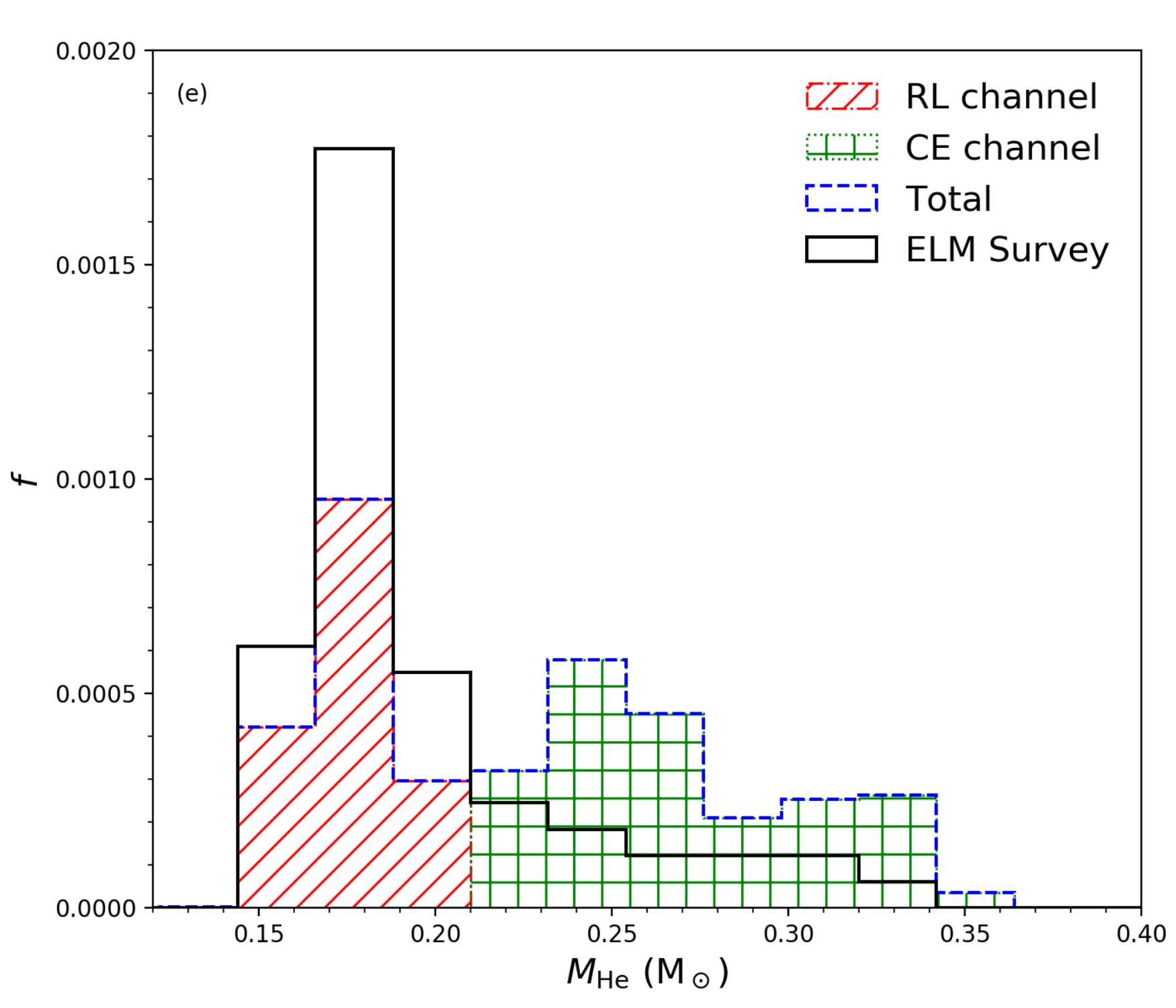}
    \caption{
	The distribution of $M_{\rm He}$ with different selection effect from our standard model. 
	The fractions of simulations are normalized to the total number of systems without any selection 
	effects, and the observations are normalized to the total number of observed samples. 
	No selection effects are considered in panel (a). In panel (b), the temperature selection effect 
	is considered, i.e. $8000<T_{\rm eff}<22000\;\rm K$. In panel (c), the selection effects of 
	$k>75 \;\rm km~ s^{-1}$ and $P_{\rm orb}<2\;\rm d$ are added. Then we put the 
	limits of $\log g$, i.e. $4.85<\log g<7.15$ in panel (d), and the magnitude limit 
	by multiplying a weight of $L_{\rm He}^{3/2}$ is included in panel (e). 
	The systems from RL channel and CE channel are shown in red and 
	green hatched regions, respectively. The ELM Survey samples are shown in solid black histogram in panel (e). 
	} 
    \label{fig:10}
\end{figure}

The mass distribution of He WD components is presented in Figure~\ref{fig:10}. The red and green 
hatched regions are for He WDs from the RL channel and CE channel, respectively, and the blue dashed line shows the 
combination of the two. No selection effects have been considered in panel (a). For panels (b) to (d), we add 
the following selection effects step by step: effective temperature in the range of $8000-22000\;\rm K$, 
the semi-amplitude $k>75\;\rm km ~s^{-1}$ and $P_{\rm orb}<2.0\;\rm d$, and the surface gravity 
$4.85<\log g<7.15$. Since brighter objects have larger probabilities of being 
detected, we simply include the magnitude limit by multiplying a weight of $L_{\rm He}^{3/2}$ in panel (e), 
as done by \citet{chen2017}, where $L_{\rm He}$ is the luminosity of He WDs. 
The fractions of simulations are normalized against the number of systems 
without any selection effects, and the observations are normalized against the total number of observed samples. 

We see that the He WD mass peaks around 0.25 $\rm M_\odot$ and 0.32 $\rm M_\odot$, resulted from the RL channel and the 
CE channel, respectively, if no selection effects have been included (panel (a)). Many products are removed by 
the constraints of effective temperature (panel (b)). The $k>75\;\rm km~s^{-1}$ and $P_{\rm orb}<2.0 \;\rm d$ 
only reduce the number from the RL channel since they are in accord with the $M_{\rm He}-P_{\rm orb}$ relation and have 
relatively long orbital period, especially when $M_{\rm He}>0.20\;\rm M_\odot$ (panel (c)). Next the constraint on 
the surface gravity removes a large fraction of He WDs with mass larger than $\sim 0.3\;\rm M_\odot$ from the CE 
channel due to their thin H-rich envelope (panel (d)). 
The fraction of low-mass He WDs becomes larger after we take the magnitude limit into consideration, 
as panel (e) shows. The reason is that most low-mass He WDs ($\lesssim 0.18\;\rm M_\odot$) are in the contraction phase 
and have high luminosity (i.e. smaller magnitude). 
However, most He WDs with $M_{\rm He}\gtrsim 0.23\;\rm M_\odot$ are located below the turnoff point 
around the large temperature end due to the much longer timescale in cooling phase in comparison to the contraction 
phase (see the discussion in Section~\ref{subsubsec:5.1.4}), and have relatively low luminosity (i.e. larger 
magnitude). As a consequence, we have two peaks, $\sim 0.18$ 
and $0.25 \;\rm M_\odot$, for the ELM WD mass distribution. All of the selection effects introduced above 
are considered in the following discussion, unless otherwise stated.

\begin{figure}
	\centering
	\includegraphics[width=0.55\textwidth]{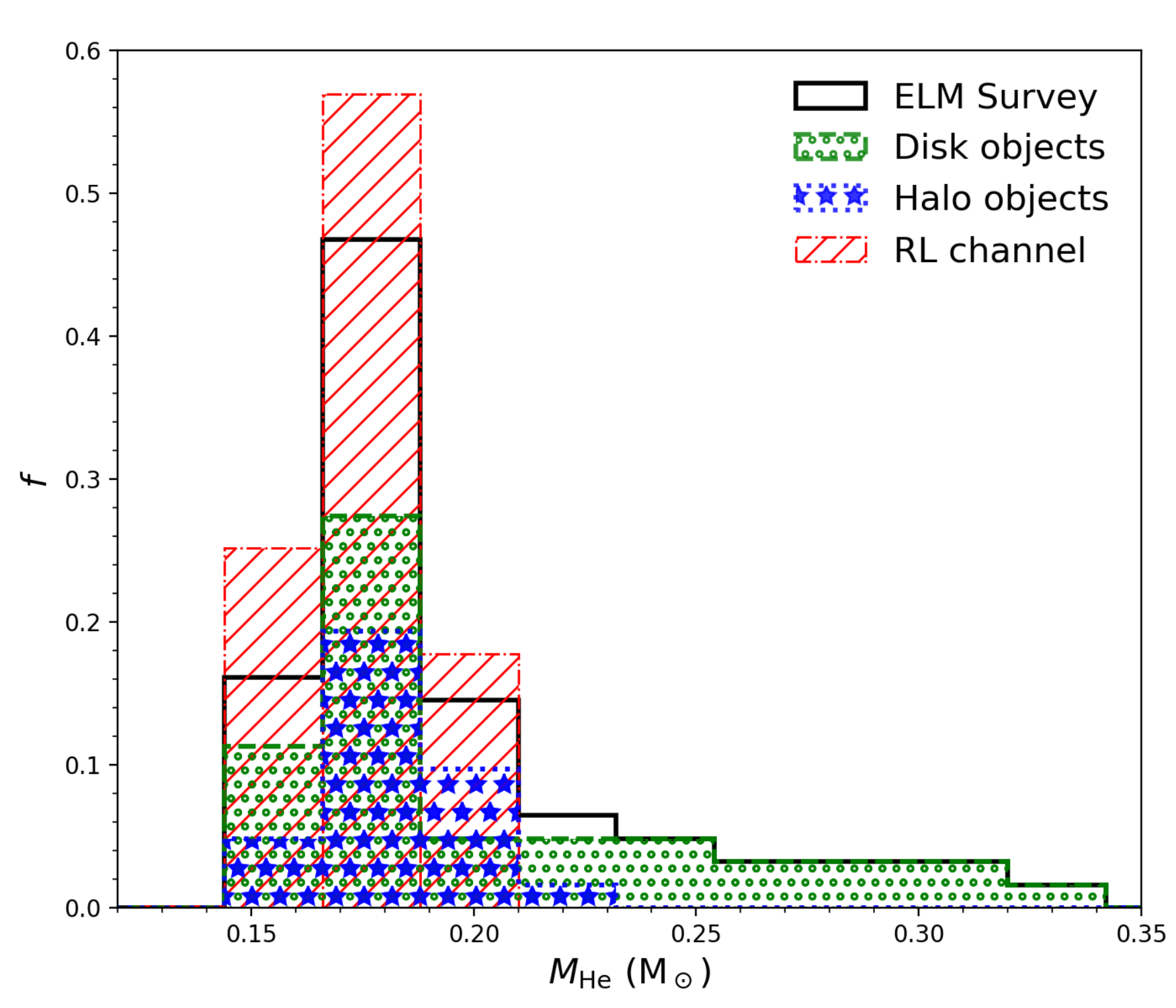}
    \caption{The comparison between mass distribution of ELM WDs from RL channel in our standard model 
	and that of observations. The green and blue regions represent the distribution of disk and 
	halo objects in ELM Survey, respectively. 
	The fractions of simulations are normalized to the total number of systems from RL channel, and the 
	observations are normalized to the respectively total number of observed samples. 
	}
    \label{fig:11}
\end{figure}

The mass distribution here is a little different from that of the observations, which has only one peak around 0.18 
$\rm M_\odot$. This suggests that the majority of the observations are from the RL channel, as shown in Figure~\ref{fig:11}, 
in which only the products from the RL channel are considered, with inclusion of selection effects mentioned 
above. It matches the observation well when $M_{\rm He}<0.22\;\rm M_\odot$. 
This mass peak comes from three factors, i.e. the parameter space for producing ELM WDs in DDs, the lifetime of low-mass 
proto-He WDs, and selection effects of $k>75\;{\rm km~s^{-1}},P_{\rm orb}<2\;\rm d$, where the selection effects are 
predominant factor (see the transition from panel (b) to panel (c) in Figure~\ref{fig:10}). 
It is a little different from that of \citet{brown2016b}, who suggested that the evolution timescale 
of low-mass proto-He WDs is the most important factor for the observed mass peak. 

For the CE channel, the predicted number is larger than the observations. 
This discrepancy may be explained by following reasons. In this work, 
we assumed that the evolutionary behaviors of ELM WDs from the CE channel are the same as those from the RL channel. 
If the envelope mass is slightly smaller than that given in this paper, the 
ELM WDs then have lower luminosity and larger surface gravity \citep{calcaferro2018}, 
and the inclusion of a magnitude limit may reduce more systems with massive He WDs. Besides, 
we checked the properties of the products with $M_{\rm He}>0.27\rm \;M_\odot$ and found that the surface gravity 
$\log g$ is larger than $\sim 7$, close to the detection limit. These systems could be removed 
if the envelope mass is smaller, because the surface gravity can be beyond the upper limit of $\log g=7.15$. 

Of course, it may also imply that many He WDs with $M_{\rm He}\gtrsim 0.22\;\rm M_\odot$ are waiting to be discovered. 

The distribution of disk and halo objects in ELM Survey is also presented separately in Figure~\ref{fig:11}. 
Almost all halo objects are less than $M_{\rm He}\lesssim0.21\;\rm M_\odot$, which suggests that the halo 
objects are mainly produced from the RL channel. Given that the metallicity in the halo ($Z\sim 0.001$) is 
generally much lower than that in the disk, we computed the formation of ELM WDs from RL channel at low 
metallicities and found that most halo objects can be explained as shown in Figure~\ref{fig:4}. 
The reason for few ELM WDs being produced from CE channel in the halo can be explained as follows.
On one hand, the binding energy of the envelope of the donor stars is larger at low metallicity. 
On the other hand, the orbital period of binaries at the onset of CE are generally smaller; hence, 
orbital energy is smaller at low metallicity. Therefore, it is difficult for these binaries to 
survive from the CE phase. Furthermore, the star formation history of the halo is significantly 
different from that of the disk. A burst of star formation was generally assumed for the halo \citep{robin2003}, 
which may also lead to a smaller number of massive He WDs in the halo at present. 
As a consequence, the ELM WDs formed from the CE channel can be very rare in the halo.
The detailed calculation of the formation of ELM WDs in the halo will be included in the next work. 

\subsubsection{Distribution of companion mass}
\label{subsubsec:5.2.2}
\begin{figure}
	\centering
	\includegraphics[width=0.55\textwidth]{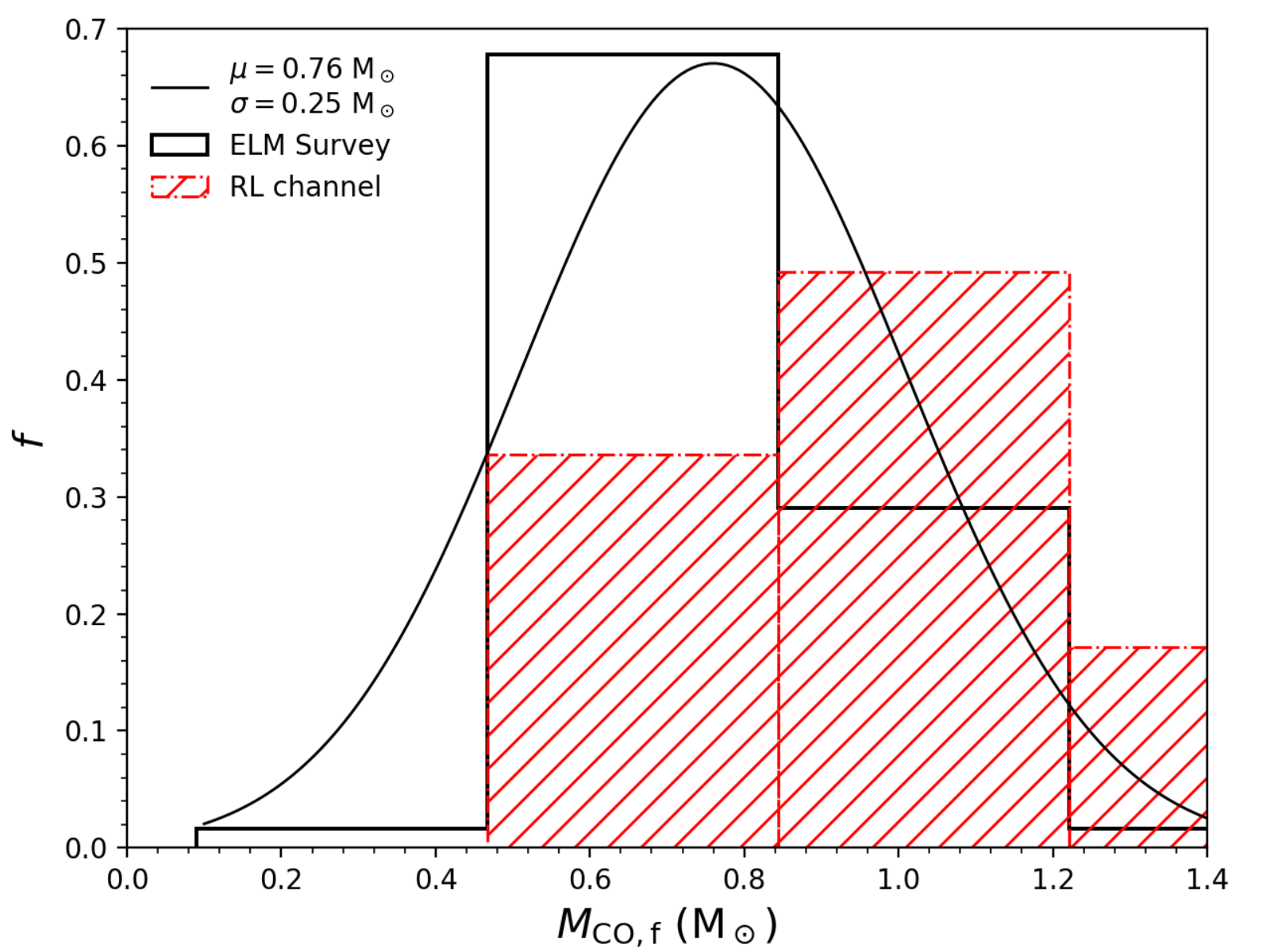}
    \caption{Comparison of CO WD mass distribution of ELM binary systems from our standard model with 
	observations. 
	The fractions of simulations are normalized to the total number of systems from RL channel, and the 
	observations are normalized to the total number of observed samples. 
	The solid black histogram is the observed samples in ELM Survey 
	and the red hatched region is our simulated results. 
	We take the mean value of error as the bin size (0.37 $\rm M_\odot$). 
	The fitted normal distribution of the observations is from \citet{brown2016a}. 
	} 
    \label{fig:12}
\end{figure}

The distribution of $M_{\rm CO,f}$ is presented in Figure~\ref{fig:12}. As we discussed above, the majority of 
the observations are from the RL channel. So, we only show the simulation results of RL channel. 
Since the error is relatively large in the observations, then we take the bin size as the mean value of 
the error, i.e. $0.37\;\rm M_\odot$. \citet{brown2016a} gave a normal distribution of CO WDs mass with 
mean value $\mu=0.76\;\rm M_\odot$ and standard deviation $\sigma=0.25\;\rm M_\odot$, which is the best match 
for the observed $k$ distribution, as the black solid line shows. 
We see that the theoretical results (most are in $0.8-1.2\;\rm M_\odot$) are larger than the observations, 
which indicates that the true accretion efficiency of CO WD may be lower than our model assumption. 

\subsubsection{Mass distribution of progenitors}
\label{subsubsec:5.2.3}
\begin{figure}
	\centering
	\includegraphics[width=0.55\textwidth]{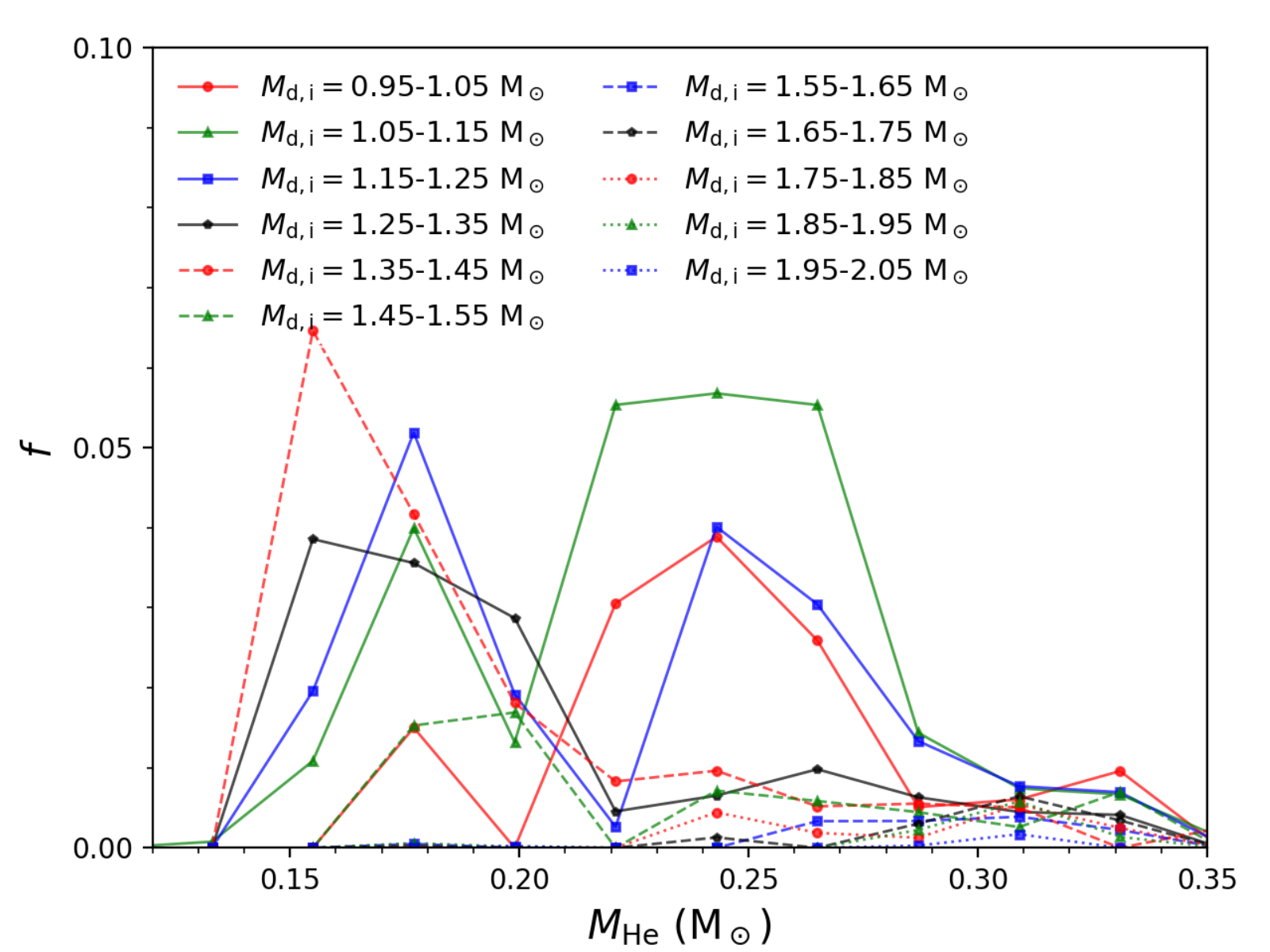}
    \caption{He WD mass distribution for different donors mass in our standard model. 
	Colors and symbols denote the initial mass range of donors. 
	The fractions are normalized to the total number of systems. 
	%The peak value is at $\sim$$0.18\;\rm M_\odot$. 
	%Proto-He WDs with mass less than $\sim0.22\;\rm M_\odot$ are mainly from the RL channel 
	%and the initial donors masses are less than $1.5\;\rm M_\odot$. While the massive 
	%proto-He WDs are produced from CE channel. 
	%The most common progenitors for low-mass He WDs ($M_{\rm He}\lesssim0.25\;\rm M_\odot$) have mass 
	%from 1.25 to 1.45$\;\rm M_\odot$. 
	}
    \label{fig:13}
\end{figure}

Figure~\ref{fig:13} shows the mass distribution of ELM WDs resulting from various progenitors in the standard model, 
where different colors and line styles are for different progenitor mass range as indicated. For each of the lines, 
the low-mass peak ($\sim 0.18\;\rm M_\odot$) is for the RL channel and the high-mass peak ($\sim 0.27\;\rm M_\odot$) 
is for the CE channel. For the RL channel, the major products come from the progenitors with mass in the 
range of $1.15-1.45\;\rm M_\odot$, since (1) the delay time for the formation of He WDs is shorter 
compared with progenitors with lower mass, so He WDs from these donors can be produced more efficiently in our 
Galaxy mode; and (2) the parameter space is larger compared with that of massive donors (see Figure~\ref{fig:5}). 
We see that the donors of mass larger than $\sim 1.6\;\rm M_\odot$ have little 
contribution. This is consistent with that of \citet{sunm2017}, who found that the maximum progenitor mass for 
ELM WDs with mass lower than $\sim 0.18\;\rm M_\odot$ is near $1.5-1.6\;\rm M_\odot$. For the CE channel, the main 
contribution comes from the progenitors with mass less than $1.25\;\rm M_\odot$. For more massive progenitors, 
the binding energy of the envelope near the base of giant branch is too high, leading the envelope to be hardly 
ejected. 

\subsubsection{The role of CE coefficient $\alpha_{\rm CE}$}
\label{subsubsec:5.2.4}
\begin{figure*}
	\centering
	\includegraphics[width=0.7\textwidth]{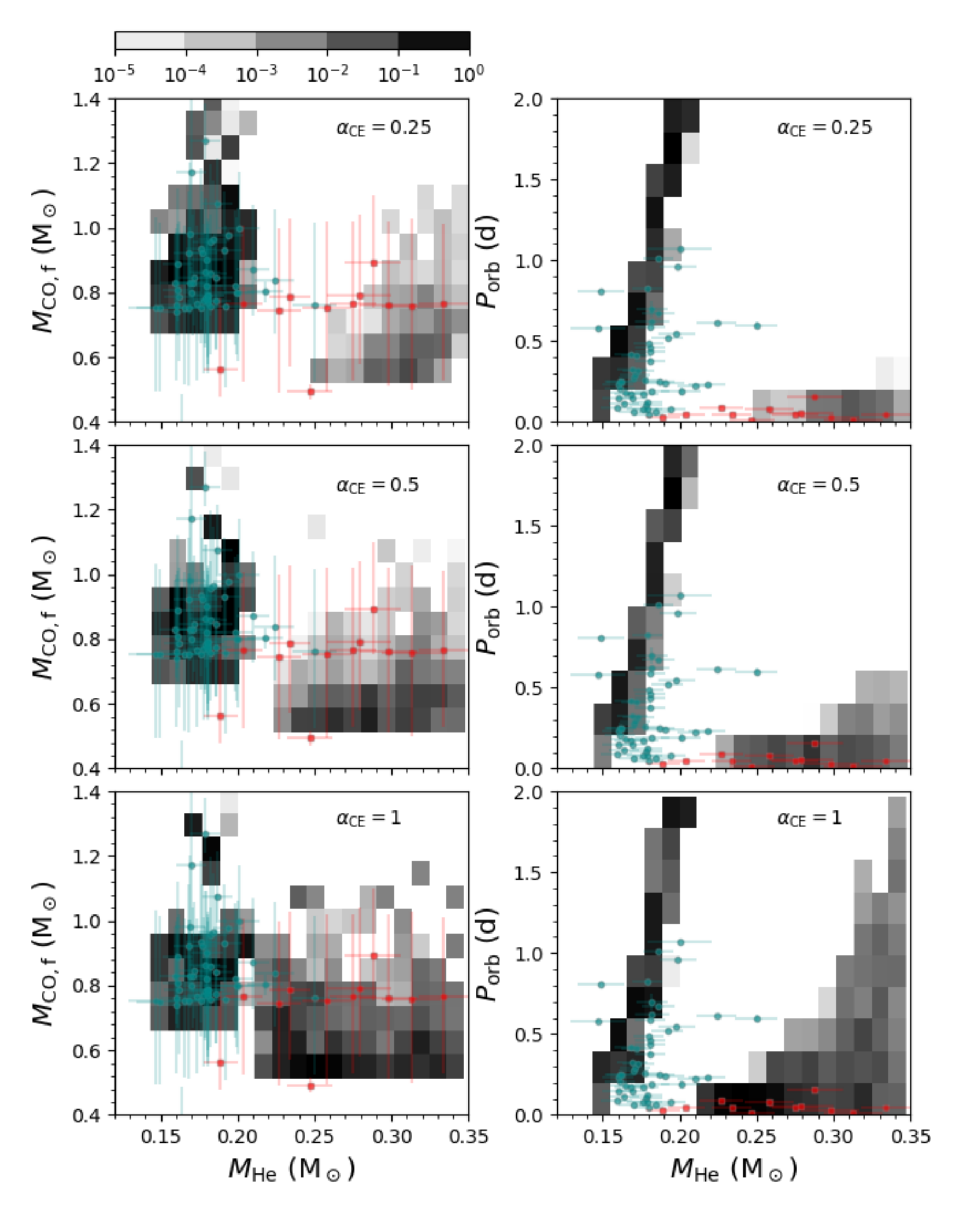}
	\caption{Binary population synthesis results with the selection effects considered 
	for different values of $\alpha_{\rm CE}$. From the first 
	row to the last row, the values of $\alpha_{\rm CE}$ are 0.25, 0.5, and 1, respectively. 
	The left column	shows the distribution of ELM binaries in the $M_{\rm He}-M_{\rm CO,f}$ plane, 
	and the right column is that in the $M_{\rm He}-P_{\rm orb}$ plane. These symbols are the 
	observed samples in the ELM Survey, which have been divided base on the formation channel, 
	and the blue circles are from the RL channel, while the red squares are from the CE channel. 
	}
    \label{fig:14}
\end{figure*}
We present the 2D distributions in the ($M_{\rm He}-M_{\rm CO,f}$) and 
($M_{\rm He}-P_{\rm orb}$) planes for different $\alpha_{\rm CE}$ in Figure~\ref{fig:14}, 
where we have distinguished different formation channels for the observed samples. 
Here the red squares represent the systems from the 
CE channel, and the blue circles represent the systems from the RL channel (see Section~\ref{subsubsec:4.2.3}). 
From the first row to the last row, the value of $\alpha_{\rm CE}$ is 0.25, 0.5, and 1, 
respectively. 
In our simulations, the systems in the area with $M_{\rm He}\lesssim\;0.22\rm M_\odot$ are mainly produced 
from the RL channel, while the other part with massive He WDs and short orbital period comes 
from the CE channel. One can easily distinguish the formation channel from the $M_{\rm He}-P_{\rm orb}$ plane. 
In the $M_{\rm CO}-M_{\rm He}$ plane, for ELM binary systems from RL channel, we can see that the peak value 
of $M_{\rm CO}$ is about $0.8-1.0\;\rm M_\odot$. And for ELM binary systems from CE channel, the peak value is 
close to $0.6\;\rm M_\odot$. Since the mass of CO WDs before the occurrence of MT peaks at about 
$\sim$$0.6\;\rm M_\odot$, if the MT process is stable, the accretors accrete materials 
to grow in mass, as discussed in Section~\ref{subsubsec:5.1.4}. However, for unstable MT process, 
the timescale of the CE phase is very short and the WDs will not increase much mass. 
The values of $\alpha_{\rm CE}$ has an important effect on the systems from CE channel. 
For larger $\alpha_{\rm CE}$, the proportion of ELM WDs from the CE channel increases, since more orbital energy 
is released to eject the CE. 
Due to more orbital energy is used for ejecting the CE. 
For the case of $\alpha_{\rm CE}=1$, the minimum mass of proto-He WDs from the CE 
channel is about $0.21\;\rm M_\odot$. This is consistent with \citet{sunm2017}, 
who used $\alpha_{\rm CE}=2$ and found that ELM WDs with mass less than $\sim$$0.18\;\rm M_\odot$ 
are hard to be formed through the CE ejection process. 

In the RL channel, relatively few ELM WDs systems with orbital periods smaller than $0.2\;\rm d$ are produced 
according to the He WD mass - orbital period plane in Figure~\ref{fig:14}. 
And the GWR merger timescale is much larger than the combined evolutionary timescale of contraction 
and cooling phases. Therefore, most of these systems within the detection limit will not become semidetached. 
However, in the CE channel, many systems have very short orbital periods ($\sim 0.01\;\rm d$) after the ejection 
of CE, and the GWR merger timescale is as low as several Myr. We find that approximately $40\%$ of systems will 
merge before going beyond the detection limit. 

In the $M_{\rm He}-P_{\rm orb}$ plane, one can notice that many observed ELM WDs from the RL channel 
have orbital periods less than $\sim 0.2\;\rm d$. However, our model for solar metallicity predicts that many 
systems from the RL channel have longer orbital periods. Two possible reasons can explain this discrepancy. 
The first one is the effect of 
metallicity. As shown in Figure~\ref{fig:4}, the orbital period for low-metallicity system is lower than 
that of Pop I system for a given He WD mass. Then we expect that more short orbital period systems will 
be produced in stellar population with low metallicity. The second 
possibility is to consider extra angular momentum loss, such as the circumbinary disk which is frequently 
used to model the evolution of cataclysmic variables (CVs, e.g. \citealt{spruit2001,taam2001,
willems2005,knigge2011}). This mechanism may reduce the final binary orbital period and partially explain 
these short-period systems. Deeper discussion is beyond the scope of this paper. 

\subsubsection{Birth rate and local space density}
\label{subsec:5.2.5}
\begin{figure}
	\centering
	\includegraphics[width=0.6\textwidth]{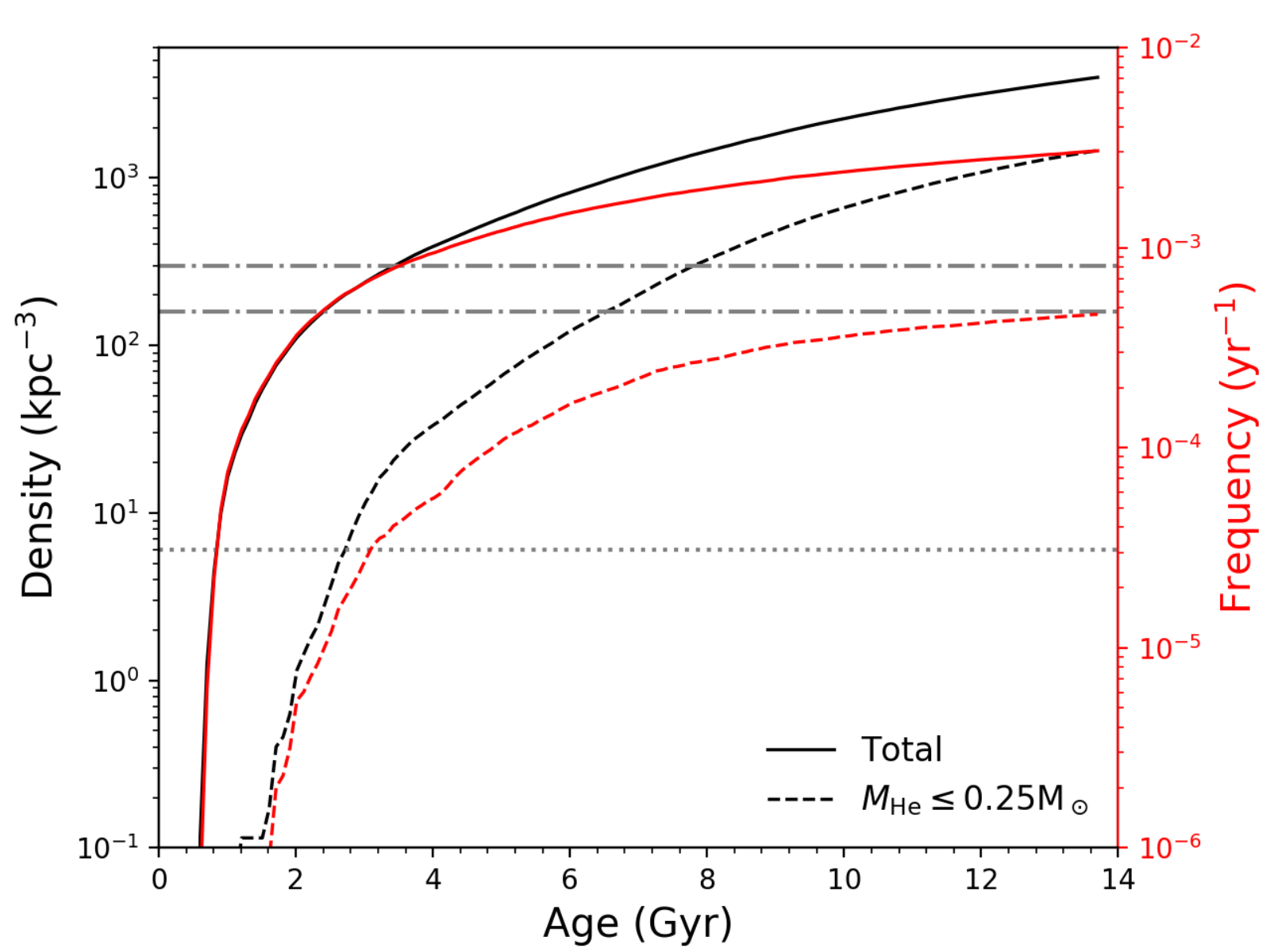}
	\caption{The local space density and birth rate of ELM binary systems for our galaxy with SFR of 
	2$\;\rm M_\odot yr^{-1}$ in our standard model. 
	The solid line represents the whole population of ELM WDs in DDs and the dashed line is 
	only for the proto-He WDs with $M_{\rm He}<0.25\;\rm M_\odot$. The local space density is shown in black 
	and the birth rate is shown in red, respectively. The two dash-dotted lines show the observed local space 
	density of disk ELM WDs, which is calculated by using different galactic model 
	\citep{juric2008,nelemans2001}. The dotted line is for the halo ELM WDs. 
	} 
    \label{fig:15}
\end{figure}

The birth rate and local space density for ELM WDs in DDs are presented in Figure~\ref{fig:15}. 
The solid lines are for the whole population of DDs containing proto-He WDs 
($M_{\rm He}\leq 0.4\;\rm M_\odot$). 
Since He WDs with a mass larger than $\gtrsim 0.25\;\rm M_\odot$ are from the CE channel in general and it is uncertain 
for the CE process as well as the products. Then, for the sake of comparison with the 
observations, we sort out the systems with $M_{\rm He}\leq0.25\;\rm M_\odot$ from 
the whole populations, as shown with the dashed lines. The volume of the galaxy is 
taken as $5\times 10^{2}\;\rm{kpc}^3$. Besides, more than $85\%$ of ELM WDs have 
mass less than $0.25\;\rm M_\odot$, so the following discussion  
only refer to these systems with ELM WDs mass $M_{\rm He}\leq 0.25\;\rm M_\odot$. 
The birth rates are shown in red lines. \citet{brown2016b} differentiated 
disk/halo objects by kinematics and estimated the local density for 
halo ELM WDs are about $160$ and $ 300\;\rm{kpc}^{-3}$ by using different disk model 
\citep{juric2008,nelemans2001}, as the dash-dotted lines show. The local space density is 
6$\;\rm{kpc}^{-3}$ for halo objects, as the dotted line shows. 
In our calculation, the local space density and birth rate are $\sim$1500 $\rm {kpc}^{-3}$ 
and $\sim$$5\times 10^{-4}\;\rm {yr}^{-1}$ at 13.7 Gyr, respectively. 
Our results are higher than the observed values, so there may remain many ELM WDs to be 
detected in the future. 

\section{Conclusions and Discussions}
\label{sec:6}
In this paper, we systematically studied the formation of ELM WDs in DDs. Both observation and theoretical 
study show that these objects can be produced either from stable MT process (RL channel) or common 
envelope ejection (CE channel). Based on detailed binary evolution, the parameter space for producing 
ELM WDs from RL channel have been shown. The basic properties of ELM WDs in DDs in this way are similar to those 
in EL CVn-type stars and those in millisecond pulsars due to similar formation processes, i.e. 
the evolution tracks, the contraction timescales from the end of mass transfer to the maximum 
temperature, H-flashes, and the ELM WD mass - orbital period relation etc.

We then further studied population properties of ELM WDs in DDs, employing a binary population 
synthesis approach. The conclusions are summarized as follows. 

\begin{itemize}
\setlength{\itemsep}{0pt}
\setlength{\parsep}{0pt}
\setlength{\parskip}{0pt}
  \item[(1)] The intrinsic mass for ELM WDs (without selection effects) peaks around $0.25\;\rm M_\odot$ 
and $0.32\;\rm M_\odot$, which are resulted from RL channel and CE channel, respectively.
  \item[(2)] If selection effects in ELM Survey are included in our models, the mass peak moves to 
$0.18\;\rm M_\odot$ for the RL channel, and to $0.25\;\rm M_\odot$ for the CE channel. Furthermore, 
the RL channel are responsible for ELM WDs in DDs with He WD mass less than $\sim0.22\;\rm M_\odot$ and 
the CE channel are for those with more massive (proto-) He WDs. 
  \item[(3)] The most likely progenitors have masses in the range of $1.15-1.45\;\rm M_\odot$ for 
the ELM WDs from RL channel, and $0.95-1.25\;\rm M_\odot$ for those from CE channel. 
  \item[(4)] The CO WD companions show two mass peaks i.e. $0.8-1.0\;\rm M_\odot$ and $\sim0.6\;\rm M_\odot$, 
which are the contribution of the RL channel and CE channel, respectively. The CO WD companions from 
the RL channel are obviously larger than that from the CE channel since the CO WDs may increase in mass 
significantly during stable MT process. 
  \item[(5)] For a constant star formation rate of $2\;\rm M_\odot{\rm{yr}}^{-1}$, the birth rate of ELM WDs in DDs 
with He WD mass less than $0.25\;\rm M_\odot$ is $\sim 5\times 10^{-4}\;\rm {yr}^{-1}$, and the local space density 
for these systems is $\sim1500\;\rm {kpc}^{-3}$, which is much higher than that of observations. 
This probably indicates that there are still many ELM WDs in DDs to be discovered in future. 
\end{itemize}
Most properties of ELM WDs in DDs from ELM Survey have been well reproduced by our theoretical study, 
especially the ELM WD mass and their CO WD companion mass from the RL channel, and the orbital period. 
However, the mass peak of ELM WDs resulted from the CE channel, $\sim0.25\;\rm M_\odot$, has not appeared in 
observation. In our study, we assumed that the ELM WDs from CE channel have the same structure as 
that from RL channel and obtained their following evolution (after the ejection of CE) by interpolating 
from the ELM WD model grid obtained from detailed binary evolution calculation. Our theoretical study 
possibly indicates that there are more ELM WDs in DDs near this mass peak awaiting for being discovered, or 
that the structure of ELM WDs from the CE channel is significantly different from that produced from the 
RL channel. Since ELM WDs in DDs from CE channel have orbital periods obviously shorter than that from 
RL channel, their contribution to the foreground of GWR is of great importance, as we show in the next paper. 
It is necessary and crucial to confirm whether the mass peak exists or not from observation. 
\section*{Acknowledgements}
We thank the anonymous referee for his/her very helpful suggestions on the manuscript. 
The authors gratefully acknowledge the computing time granted by the 
Yunnan Observatories, and provided on the facilities at the Yunnan 
Observatories Supercomputing Platform. 
This work is partially supported by the Natural Science Foundation of China (Grant no. 
11733008, 11521303, 11703081, 11422324), by the National Ten-thousand talents program, 
by Yunnan province (No. 2017HC018), by Youth Innovation Promotion Association of the 
Chinese Academy of Sciences and the CAS light of West China Program. 
\software{BSE \citep{hurley00,hurley02}}
\software{MESA (v9575; \citealt{paxton2011,paxton2013,paxton2015})}
%%%%%%%%%%%%%%%%%%%%%%%%%%%%%%%%%%%%%%%%%%%%%%%%%%



\bibliographystyle{./aasjournal}
\bibliography{./elmwd}

\begin{thebibliography}{}
\expandafter\ifx\csname natexlab\endcsname\relax\def\natexlab#1{#1}\fi
\providecommand{\url}[1]{\href{#1}{#1}}
\providecommand{\dodoi}[1]{doi:~\href{http://doi.org/#1}{\nolinkurl{#1}}}
\providecommand{\doeprint}[1]{\href{http://ascl.net/#1}{\nolinkurl{http://ascl.net/#1}}}
\providecommand{\doarXiv}[1]{\href{https://arxiv.org/abs/#1}{\nolinkurl{https://arxiv.org/abs/#1}}}

\bibitem[{{Amaro-Seoane} {et~al.}(2012){Amaro-Seoane}, {Aoudia}, {Babak},
  {Bin{\'e}truy}, {Berti}, {Boh{\'e}}, {Caprini}, {Colpi}, {Cornish},
  {Danzmann}, {Dufaux}, {Gair}, {Jennrich}, {Jetzer}, {Klein}, {Lang}, {Lobo},
  {Littenberg}, {McWilliams}, {Nelemans}, {Petiteau}, {Porter}, {Schutz},
  {Sesana}, {Stebbins}, {Sumner}, {Vallisneri}, {Vitale}, {Volonteri}, \&
  {Ward}}]{lisa2012}
{Amaro-Seoane}, P., {Aoudia}, S., {Babak}, S., {et~al.} 2012, Classical and
  Quantum Gravity, 29, 124016, \dodoi{10.1088/0264-9381/29/12/124016}

\bibitem[{{Breton} {et~al.}(2012){Breton}, {Rappaport}, {van Kerkwijk}, \&
  {Carter}}]{breton2012}
{Breton}, R.~P., {Rappaport}, S.~A., {van Kerkwijk}, M.~H., \& {Carter}, J.~A.
  2012, \apj, 748, 115, \dodoi{10.1088/0004-637X/748/2/115}

\bibitem[{{Brown} {et~al.}(2016{\natexlab{a}}){Brown}, {Gianninas}, {Kilic},
  {Kenyon}, \& {Allende Prieto}}]{brown2016a}
{Brown}, W.~R., {Gianninas}, A., {Kilic}, M., {Kenyon}, S.~J., \& {Allende
  Prieto}, C. 2016{\natexlab{a}}, \apj, 818, 155,
  \dodoi{10.3847/0004-637X/818/2/155}

\bibitem[{{Brown} {et~al.}(2013){Brown}, {Kilic}, {Allende Prieto},
  {Gianninas}, \& {Kenyon}}]{brown2013}
{Brown}, W.~R., {Kilic}, M., {Allende Prieto}, C., {Gianninas}, A., \&
  {Kenyon}, S.~J. 2013, \apj, 769, 66, \dodoi{10.1088/0004-637X/769/1/66}

\bibitem[{{Brown} {et~al.}(2010){Brown}, {Kilic}, {Allende Prieto}, \&
  {Kenyon}}]{brown2010}
{Brown}, W.~R., {Kilic}, M., {Allende Prieto}, C., \& {Kenyon}, S.~J. 2010,
  \apj, 723, 1072, \dodoi{10.1088/0004-637X/723/2/1072}

\bibitem[{{Brown} {et~al.}(2012){Brown}, {Kilic}, {Allende Prieto}, \&
  {Kenyon}}]{brown2012}
---. 2012, \apj, 744, 142, \dodoi{10.1088/0004-637X/744/2/142}

\bibitem[{{Brown} {et~al.}(2011){Brown}, {Kilic}, {Hermes}, {Allende Prieto},
  {Kenyon}, \& {Winget}}]{brown2011}
{Brown}, W.~R., {Kilic}, M., {Hermes}, J.~J., {et~al.} 2011, \apjl, 737, L23,
  \dodoi{10.1088/2041-8205/737/1/L23}

\bibitem[{{Brown} {et~al.}(2016{\natexlab{b}}){Brown}, {Kilic}, {Kenyon}, \&
  {Gianninas}}]{brown2016b}
{Brown}, W.~R., {Kilic}, M., {Kenyon}, S.~J., \& {Gianninas}, A.
  2016{\natexlab{b}}, \apj, 824, 46, \dodoi{10.3847/0004-637X/824/1/46}

\bibitem[{{Brown} {et~al.}(2017){Brown}, {Kilic}, {Kosakowski}, \&
  {Gianninas}}]{brown2017}
{Brown}, W.~R., {Kilic}, M., {Kosakowski}, A., \& {Gianninas}, A. 2017, \apj,
  847, 10, \dodoi{10.3847/1538-4357/aa8724}

\bibitem[{{Calcaferro} {et~al.}(2018){Calcaferro}, {Althaus}, \&
  {C{\'o}rsico}}]{calcaferro2018}
{Calcaferro}, L.~M., {Althaus}, L.~G., \& {C{\'o}rsico}, A.~H. 2018, ArXiv
  e-prints.
\newblock \doarXiv{1802.06753}

\bibitem[{{Calcaferro} {et~al.}(2017){Calcaferro}, {C{\'o}rsico}, \&
  {Althaus}}]{calcaferro2017}
{Calcaferro}, L.~M., {C{\'o}rsico}, A.~H., \& {Althaus}, L.~G. 2017, \aap, 607,
  A33, \dodoi{10.1051/0004-6361/201731230}

\bibitem[{{Carter} {et~al.}(2011){Carter}, {Rappaport}, \&
  {Fabrycky}}]{carter2011}
{Carter}, J.~A., {Rappaport}, S., \& {Fabrycky}, D. 2011, \apj, 728, 139,
  \dodoi{10.1088/0004-637X/728/2/139}

\bibitem[{{Chen} \& {Han}(2002)}]{chen2002}
{Chen}, X., \& {Han}, Z. 2002, \mnras, 335, 948,
  \dodoi{10.1046/j.1365-8711.2002.05680.x}

\bibitem[{{Chen} \& {Han}(2003)}]{chen2003}
---. 2003, \mnras, 341, 662, \dodoi{10.1046/j.1365-8711.2003.06449.x}

\bibitem[{{Chen} {et~al.}(2017){Chen}, {Maxted}, {Li}, \& {Han}}]{chen2017}
{Chen}, X., {Maxted}, P.~F.~L., {Li}, J., \& {Han}, Z. 2017, \mnras, 467, 1874,
  \dodoi{10.1093/mnras/stx115}

\bibitem[{{Chomiuk} \& {Povich}(2011)}]{chomiuk2011}
{Chomiuk}, L., \& {Povich}, M.~S. 2011, \aj, 142, 197,
  \dodoi{10.1088/0004-6256/142/6/197}

\bibitem[{{C{\'o}rsico} \& {Althaus}(2014{\natexlab{a}})}]{corsico2014a}
{C{\'o}rsico}, A.~H., \& {Althaus}, L.~G. 2014{\natexlab{a}}, \aap, 569, A106,
  \dodoi{10.1051/0004-6361/201424352}

\bibitem[{{C{\'o}rsico} \& {Althaus}(2014{\natexlab{b}})}]{corsico2014b}
---. 2014{\natexlab{b}}, \apjl, 793, L17, \dodoi{10.1088/2041-8205/793/1/L17}

\bibitem[{{C{\'o}rsico} {et~al.}(2016){C{\'o}rsico}, {Althaus}, {Serenelli},
  {Kepler}, {Jeffery}, \& {Corti}}]{corsico2016}
{C{\'o}rsico}, A.~H., {Althaus}, L.~G., {Serenelli}, A.~M., {et~al.} 2016,
  \aap, 588, A74, \dodoi{10.1051/0004-6361/201528032}

\bibitem[{{C{\'o}rsico} {et~al.}(2012){C{\'o}rsico}, {Romero}, {Althaus}, \&
  {Hermes}}]{corsico2012}
{C{\'o}rsico}, A.~H., {Romero}, A.~D., {Althaus}, L.~G., \& {Hermes}, J.~J.
  2012, \aap, 547, A96, \dodoi{10.1051/0004-6361/201220114}

\bibitem[{{De Kool}(1990)}]{dekool1990}
{De Kool}, M. 1990, \apj, 358, 189, \dodoi{10.1086/168974}

\bibitem[{{Dewi} \& {Tauris}(2000)}]{dewi2000}
{Dewi}, J.~D.~M., \& {Tauris}, T.~M. 2000, \aap, 360, 1043

\bibitem[{{Eggleton}(1983)}]{eggleton1983}
{Eggleton}, P.~P. 1983, \apj, 268, 368, \dodoi{10.1086/160960}

\bibitem[{{Eggleton} {et~al.}(1989){Eggleton}, {Fitchett}, \&
  {Tout}}]{eggleton1989}
{Eggleton}, P.~P., {Fitchett}, M.~J., \& {Tout}, C.~A. 1989, \apj, 347, 998,
  \dodoi{10.1086/168190}

\bibitem[{{Gianninas} {et~al.}(2016){Gianninas}, {Curd}, {Fontaine}, {Brown},
  \& {Kilic}}]{gianninas2016}
{Gianninas}, A., {Curd}, B., {Fontaine}, G., {Brown}, W.~R., \& {Kilic}, M.
  2016, \apjl, 822, L27, \dodoi{10.3847/2041-8205/822/2/L27}

\bibitem[{{Gianninas} {et~al.}(2015){Gianninas}, {Kilic}, {Brown}, {Canton}, \&
  {Kenyon}}]{gianninas2015}
{Gianninas}, A., {Kilic}, M., {Brown}, W.~R., {Canton}, P., \& {Kenyon}, S.~J.
  2015, \apj, 812, 167, \dodoi{10.1088/0004-637X/812/2/167}

\bibitem[{{Hachisu} {et~al.}(1996){Hachisu}, {Kato}, \& {Nomoto}}]{hachisu1996}
{Hachisu}, I., {Kato}, M., \& {Nomoto}, K. 1996, \apjl, 470, L97,
  \dodoi{10.1086/310303}

\bibitem[{{Hachisu} {et~al.}(1999){Hachisu}, {Kato}, \& {Nomoto}}]{hachisu1999}
---. 1999, \apj, 522, 487, \dodoi{10.1086/307608}

\bibitem[{{Han}(1998)}]{han1998}
{Han}, Z. 1998, \mnras, 296, 1019, \dodoi{10.1046/j.1365-8711.1998.01475.x}

\bibitem[{{Han} \& {Podsiadlowski}(2004)}]{han2004}
{Han}, Z., \& {Podsiadlowski}, P. 2004, \mnras, 350, 1301,
  \dodoi{10.1111/j.1365-2966.2004.07713.x}

\bibitem[{{Han} {et~al.}(1994){Han}, {Podsiadlowski}, \& {Eggleton}}]{han1994}
{Han}, Z., {Podsiadlowski}, P., \& {Eggleton}, P.~P. 1994, \mnras, 270, 121,
  \dodoi{10.1093/mnras/270.1.121}

\bibitem[{{Han} {et~al.}(2000){Han}, {Tout}, \& {Eggleton}}]{han2000}
{Han}, Z., {Tout}, C.~A., \& {Eggleton}, P.~P. 2000, \mnras, 319, 215,
  \dodoi{10.1046/j.1365-8711.2000.03839.x}

\bibitem[{{Hermes} {et~al.}(2013){Hermes}, {Montgomery}, {Gianninas}, {Winget},
  {Brown}, {Harrold}, {Bell}, {Kenyon}, {Kilic}, \& {Castanheira}}]{hermes2013}
{Hermes}, J.~J., {Montgomery}, M.~H., {Gianninas}, A., {et~al.} 2013, \mnras,
  436, 3573, \dodoi{10.1093/mnras/stt1835}

\bibitem[{{Hurley} {et~al.}(2000){Hurley}, {Pols}, \& {Tout}}]{hurley00}
{Hurley}, J.~R., {Pols}, O.~R., \& {Tout}, C.~A. 2000, \mnras, 315, 543,
  \dodoi{10.1046/j.1365-8711.2000.03426.x}

\bibitem[{{Hurley} {et~al.}(2002){Hurley}, {Tout}, \& {Pols}}]{hurley02}
{Hurley}, J.~R., {Tout}, C.~A., \& {Pols}, O.~R. 2002, \mnras, 329, 897,
  \dodoi{10.1046/j.1365-8711.2002.05038.x}

\bibitem[{{Istrate} {et~al.}(2016{\natexlab{a}}){Istrate}, {Fontaine},
  {Gianninas}, {Grassitelli}, {Marchant}, {Tauris}, \& {Langer}}]{istrate2016b}
{Istrate}, A.~G., {Fontaine}, G., {Gianninas}, A., {et~al.} 2016{\natexlab{a}},
  \aap, 595, L12, \dodoi{10.1051/0004-6361/201629876}

\bibitem[{{Istrate} {et~al.}(2016{\natexlab{b}}){Istrate}, {Marchant},
  {Tauris}, {Langer}, {Stancliffe}, \& {Grassitelli}}]{istrate2016a}
{Istrate}, A.~G., {Marchant}, P., {Tauris}, T.~M., {et~al.} 2016{\natexlab{b}},
  \aap, 595, A35, \dodoi{10.1051/0004-6361/201628874}

\bibitem[{{Istrate} {et~al.}(2014{\natexlab{a}}){Istrate}, {Tauris}, \&
  {Langer}}]{istrate2014a}
{Istrate}, A.~G., {Tauris}, T.~M., \& {Langer}, N. 2014{\natexlab{a}}, \aap,
  571, A45, \dodoi{10.1051/0004-6361/201424680}

\bibitem[{{Istrate} {et~al.}(2014{\natexlab{b}}){Istrate}, {Tauris}, {Langer},
  \& {Antoniadis}}]{istrate2014b}
{Istrate}, A.~G., {Tauris}, T.~M., {Langer}, N., \& {Antoniadis}, J.
  2014{\natexlab{b}}, \aap, 571, L3, \dodoi{10.1051/0004-6361/201424681}

\bibitem[{{Jeffery} \& {Saio}(2013)}]{jeffery2013}
{Jeffery}, C.~S., \& {Saio}, H. 2013, \mnras, 435, 885,
  \dodoi{10.1093/mnras/stt1360}

\bibitem[{{Jia} \& {Li}(2014)}]{jia2014}
{Jia}, K., \& {Li}, X.-D. 2014, \apj, 791, 127,
  \dodoi{10.1088/0004-637X/791/2/127}

\bibitem[{{Juri{\'c}} {et~al.}(2008){Juri{\'c}}, {Ivezi{\'c}}, {Brooks},
  {Lupton}, {Schlegel}, {Finkbeiner}, {Padmanabhan}, {Bond}, {Sesar},
  {Rockosi}, {Knapp}, {Gunn}, {Sumi}, {Schneider}, {Barentine}, {Brewington},
  {Brinkmann}, {Fukugita}, {Harvanek}, {Kleinman}, {Krzesinski}, {Long},
  {Neilsen}, {Nitta}, {Snedden}, \& {York}}]{juric2008}
{Juri{\'c}}, M., {Ivezi{\'c}}, {\v Z}., {Brooks}, A., {et~al.} 2008, \apj, 673,
  864, \dodoi{10.1086/523619}

\bibitem[{{Kato} \& {Hachisu}(2004)}]{kato2004}
{Kato}, M., \& {Hachisu}, I. 2004, \apjl, 613, L129, \dodoi{10.1086/425249}

\bibitem[{{Kilic} {et~al.}(2011){Kilic}, {Brown}, {Allende Prieto},
  {Ag{\"u}eros}, {Heinke}, \& {Kenyon}}]{kilic2011a}
{Kilic}, M., {Brown}, W.~R., {Allende Prieto}, C., {et~al.} 2011, \apj, 727, 3,
  \dodoi{10.1088/0004-637X/727/1/3}

\bibitem[{{Kilic} {et~al.}(2012){Kilic}, {Brown}, {Allende Prieto}, {Kenyon},
  {Heinke}, {Ag{\"u}eros}, \& {Kleinman}}]{kilic2012}
---. 2012, \apj, 751, 141, \dodoi{10.1088/0004-637X/751/2/141}

\bibitem[{{Knigge} {et~al.}(2011){Knigge}, {Baraffe}, \&
  {Patterson}}]{knigge2011}
{Knigge}, C., {Baraffe}, I., \& {Patterson}, J. 2011, \apjs, 194, 28,
  \dodoi{10.1088/0067-0049/194/2/28}

\bibitem[{{Kroupa} {et~al.}(1993){Kroupa}, {Tout}, \& {Gilmore}}]{kroupa1993}
{Kroupa}, P., {Tout}, C.~A., \& {Gilmore}, G. 1993, \mnras, 262, 545,
  \dodoi{10.1093/mnras/262.3.545}

\bibitem[{{Landau} \& {Lifshitz}(1975)}]{landau}
{Landau}, L.~D., \& {Lifshitz}, E.~M. 1975, {The classical theory of fields}
  (New York: Pergamon Press,~Oxford)

\bibitem[{{Lin} {et~al.}(2011){Lin}, {Rappaport}, {Podsiadlowski}, {Nelson},
  {Paxton}, \& {Todorov}}]{lin2011}
{Lin}, J., {Rappaport}, S., {Podsiadlowski}, P., {et~al.} 2011, \apj, 732, 70,
  \dodoi{10.1088/0004-637X/732/2/70}

\bibitem[{{Livio} \& {Soker}(1988)}]{livio1988}
{Livio}, M., \& {Soker}, N. 1988, \apj, 329, 764, \dodoi{10.1086/166419}

\bibitem[{{Luo} {et~al.}(2016){Luo}, {Chen}, {Duan}, {Gong}, {Hu}, {Ji}, {Liu},
  {Mei}, {Milyukov}, {Sazhin}, {Shao}, {Toth}, {Tu}, {Wang}, {Wang}, {Yeh},
  {Zhan}, {Zhang}, {Zharov}, \& {Zhou}}]{luo2015}
{Luo}, J., {Chen}, L.-S., {Duan}, H.-Z., {et~al.} 2016, Classical and Quantum
  Gravity, 33, 035010, \dodoi{10.1088/0264-9381/33/3/035010}

\bibitem[{{Maxted} {et~al.}(2014{\natexlab{a}}){Maxted}, {Serenelli}, {Marsh},
  {Catal{\'a}n}, {Mahtani}, \& {Dhillon}}]{maxted2014b}
{Maxted}, P.~F.~L., {Serenelli}, A.~M., {Marsh}, T.~R., {et~al.}
  2014{\natexlab{a}}, \mnras, 444, 208, \dodoi{10.1093/mnras/stu1465}

\bibitem[{{Maxted} {et~al.}(2011){Maxted}, {Anderson}, {Burleigh}, {Collier
  Cameron}, {Heber}, {G{\"a}nsicke}, {Geier}, {Kupfer}, {Marsh}, {Nelemans},
  {O'Toole}, {{\O}stensen}, {Smalley}, \& {West}}]{maxted2011}
{Maxted}, P.~F.~L., {Anderson}, D.~R., {Burleigh}, M.~R., {et~al.} 2011,
  \mnras, 418, 1156, \dodoi{10.1111/j.1365-2966.2011.19567.x}

\bibitem[{{Maxted} {et~al.}(2013){Maxted}, {Serenelli}, {Miglio}, {Marsh},
  {Heber}, {Dhillon}, {Littlefair}, {Copperwheat}, {Smalley}, {Breedt}, \&
  {Schaffenroth}}]{maxted2013}
{Maxted}, P.~F.~L., {Serenelli}, A.~M., {Miglio}, A., {et~al.} 2013, \nat, 498,
  463, \dodoi{10.1038/nature12192}

\bibitem[{{Maxted} {et~al.}(2014{\natexlab{b}}){Maxted}, {Bloemen}, {Heber},
  {Geier}, {Wheatley}, {Marsh}, {Breedt}, {Sebastian}, {Faillace}, {Owen},
  {Pulley}, {Smith}, {Kolb}, {Haswell}, {Southworth}, {Anderson}, {Smalley},
  {Collier Cameron}, {Hebb}, {Simpson}, {West}, {Bochinski}, {Busuttil}, \&
  {Hadigal}}]{maxted2014}
{Maxted}, P.~F.~L., {Bloemen}, S., {Heber}, U., {et~al.} 2014{\natexlab{b}},
  \mnras, 437, 1681, \dodoi{10.1093/mnras/stt2007}

\bibitem[{{Mazeh} {et~al.}(1992){Mazeh}, {Goldberg}, {Duquennoy}, \&
  {Mayor}}]{mazeh1992}
{Mazeh}, T., {Goldberg}, D., {Duquennoy}, A., \& {Mayor}, M. 1992, \apj, 401,
  265, \dodoi{10.1086/172058}

\bibitem[{{Miller} \& {Scalo}(1979)}]{Miller1979}
{Miller}, G.~E., \& {Scalo}, J.~M. 1979, \apjs, 41, 513, \dodoi{10.1086/190629}

\bibitem[{{Nelemans} {et~al.}(2001){Nelemans}, {Portegies Zwart}, {Verbunt}, \&
  {Yungelson}}]{nelemans2001}
{Nelemans}, G., {Portegies Zwart}, S.~F., {Verbunt}, F., \& {Yungelson}, L.~R.
  2001, \aap, 368, 939, \dodoi{10.1051/0004-6361:20010049}

\bibitem[{{Nelson} {et~al.}(2004){Nelson}, {Dubeau}, \&
  {MacCannell}}]{nelson2004}
{Nelson}, L.~A., {Dubeau}, E., \& {MacCannell}, K.~A. 2004, \apj, 616, 1124,
  \dodoi{10.1086/421698}

\bibitem[{{Nomoto} {et~al.}(2007){Nomoto}, {Saio}, {Kato}, \&
  {Hachisu}}]{nomoto2007}
{Nomoto}, K., {Saio}, H., {Kato}, M., \& {Hachisu}, I. 2007, \apj, 663, 1269,
  \dodoi{10.1086/518465}

\bibitem[{{Paxton} {et~al.}(2011){Paxton}, {Bildsten}, {Dotter}, {Herwig},
  {Lesaffre}, \& {Timmes}}]{paxton2011}
{Paxton}, B., {Bildsten}, L., {Dotter}, A., {et~al.} 2011, \apjs, 192, 3,
  \dodoi{10.1088/0067-0049/192/1/3}

\bibitem[{{Paxton} {et~al.}(2013){Paxton}, {Cantiello}, {Arras}, {Bildsten},
  {Brown}, {Dotter}, {Mankovich}, {Montgomery}, {Stello}, {Timmes}, \&
  {Townsend}}]{paxton2013}
{Paxton}, B., {Cantiello}, M., {Arras}, P., {et~al.} 2013, \apjs, 208, 4,
  \dodoi{10.1088/0067-0049/208/1/4}

\bibitem[{{Paxton} {et~al.}(2015){Paxton}, {Marchant}, {Schwab}, {Bauer},
  {Bildsten}, {Cantiello}, {Dessart}, {Farmer}, {Hu}, {Langer}, {Townsend},
  {Townsley}, \& {Timmes}}]{paxton2015}
{Paxton}, B., {Marchant}, P., {Schwab}, J., {et~al.} 2015, \apjs, 220, 15,
  \dodoi{10.1088/0067-0049/220/1/15}

\bibitem[{{Rappaport} {et~al.}(2015){Rappaport}, {Nelson}, {Levine},
  {Sanchis-Ojeda}, {Gandolfi}, {Nowak}, {Palle}, \& {Prsa}}]{rappaport2015}
{Rappaport}, S., {Nelson}, L., {Levine}, A., {et~al.} 2015, \apj, 803, 82,
  \dodoi{10.1088/0004-637X/803/2/82}

\bibitem[{{Rappaport} {et~al.}(1983){Rappaport}, {Verbunt}, \&
  {Joss}}]{rappaport1983}
{Rappaport}, S., {Verbunt}, F., \& {Joss}, P.~C. 1983, \apj, 275, 713,
  \dodoi{10.1086/161569}

\bibitem[{{Ritter}(1988)}]{ritter1988}
{Ritter}, H. 1988, \aap, 202, 93

\bibitem[{{Robin} {et~al.}(2003){Robin}, {Reyl{\'e}}, {Derri{\`e}re}, \&
  {Picaud}}]{robin2003}
{Robin}, A.~C., {Reyl{\'e}}, C., {Derri{\`e}re}, S., \& {Picaud}, S. 2003,
  \aap, 409, 523, \dodoi{10.1051/0004-6361:20031117}

\bibitem[{{Spruit} \& {Taam}(2001)}]{spruit2001}
{Spruit}, H.~C., \& {Taam}, R.~E. 2001, \apj, 548, 900, \dodoi{10.1086/319030}

\bibitem[{{Sun} \& {Arras}(2017)}]{sunm2017}
{Sun}, M., \& {Arras}, P. 2017, ArXiv e-prints.
\newblock \doarXiv{1703.01648}

\bibitem[{{Taam} \& {Spruit}(2001)}]{taam2001}
{Taam}, R.~E., \& {Spruit}, H.~C. 2001, \apj, 561, 329, \dodoi{10.1086/322331}

\bibitem[{{Tauris} \& {Savonije}(1999)}]{tauris1999}
{Tauris}, T.~M., \& {Savonije}, G.~J. 1999, \aap, 350, 928

\bibitem[{{Van Grootel} {et~al.}(2013){Van Grootel}, {Fontaine}, {Brassard}, \&
  {Dupret}}]{grootel2013}
{Van Grootel}, V., {Fontaine}, G., {Brassard}, P., \& {Dupret}, M.-A. 2013,
  \apj, 762, 57, \dodoi{10.1088/0004-637X/762/1/57}

\bibitem[{{van Kerkwijk} {et~al.}(2010){van Kerkwijk}, {Rappaport}, {Breton},
  {Justham}, {Podsiadlowski}, \& {Han}}]{kerkwijk2010}
{van Kerkwijk}, M.~H., {Rappaport}, S.~A., {Breton}, R.~P., {et~al.} 2010,
  \apj, 715, 51, \dodoi{10.1088/0004-637X/715/1/51}

\bibitem[{{Webbink}(1975)}]{webbink1975}
{Webbink}, R.~F. 1975, \mnras, 171, 555, \dodoi{10.1093/mnras/171.3.555}

\bibitem[{{Webbink}(1984)}]{webbink1984}
---. 1984, \apj, 277, 355, \dodoi{10.1086/161701}

\bibitem[{{Willems} \& {Kolb}(2004)}]{willems2004}
{Willems}, B., \& {Kolb}, U. 2004, \aap, 419, 1057,
  \dodoi{10.1051/0004-6361:20040085}

\bibitem[{{Willems} {et~al.}(2005){Willems}, {Kolb}, {Sandquist}, {Taam}, \&
  {Dubus}}]{willems2005}
{Willems}, B., {Kolb}, U., {Sandquist}, E.~L., {Taam}, R.~E., \& {Dubus}, G.
  2005, \apj, 635, 1263, \dodoi{10.1086/498010}

\bibitem[{{Zhang} {et~al.}(2016){Zhang}, {Fu}, {Li}, {Ren}, \&
  {Luo}}]{zhangx2016}
{Zhang}, X.~B., {Fu}, J.~N., {Li}, Y., {Ren}, A.~B., \& {Luo}, C.~Q. 2016,
  \apjl, 821, L32, \dodoi{10.3847/2041-8205/821/2/L32}

\bibitem[{{Zorotovic} {et~al.}(2010){Zorotovic}, {Schreiber}, {G{\"a}nsicke},
  \& {Nebot G{\'o}mez-Mor{\'a}n}}]{zorotovic2010}
{Zorotovic}, M., {Schreiber}, M.~R., {G{\"a}nsicke}, B.~T., \& {Nebot
  G{\'o}mez-Mor{\'a}n}, A. 2010, \aap, 520, A86,
  \dodoi{10.1051/0004-6361/200913658}

\end{thebibliography}

\include{table_information}
\include{table_KS}
\include{figure}

\end{document}